\documentclass[pdflatex,sn-nature]{sn-jnl}

\usepackage{graphicx}
\usepackage{amsmath,amssymb,amsfonts}
\usepackage{bm,mathtools}
\usepackage{siunitx}
\usepackage{subcaption}
\usepackage{placeins}
\usepackage{xurl}

\begin{document}

\title[Integration and inference in neuronal cultures]{Bridging integrated information theory and the free-energy principle in living neuronal networks}

\author[1]{\fnm{Teruki} \sur{Mayama}}
\author[1]{\fnm{Dai} \sur{Akita}}
\author[1]{\fnm{Sota} \sur{Shimizu}}
\author[1]{\fnm{Yuki} \sur{Takano}}
\author*[1,2]{\fnm{Hirokazu} \sur{Takahashi}}\email{takahashi@i.u-tokyo.ac.jp}

\affil[1]{\orgdiv{Department of Mechano-Informatics, Graduate School of Information Science and Technology}, \orgname{The University of Tokyo}, \orgaddress{\city{Tokyo}, \country{Japan}}}
\affil[2]{\orgdiv{International Research Center for Neurointelligence (WPI-IRCN), The University of Tokyo Institutes for Advanced Study (UTIAS)}, \orgname{The University of Tokyo}, \orgaddress{\city{Tokyo}, \country{Japan}}}

\abstract{Integrated Information Theory (IIT) links consciousness to integrated causal structure, whereas the Free-Energy Principle (FEP) explains self-organization through variational free-energy minimization. Their relationship in living neural systems remains unclear. We analyzed dissociated neuronal cultures learning to infer hidden signal sources. Across repeated stimulation, variational free energy decreased, while inference accuracy and Bayesian surprise—the divergence between prior and posterior beliefs—increased. An IIT-inspired integrated-information proxy and main-complex size followed a non-monotonic, hill-shaped trajectory. The proxy correlated most strongly with Bayesian surprise and more weakly with accuracy and variational free energy. An Ising-model analysis indicated that Bayesian surprise and integrated information can be jointly amplified near shared positive critical modes and suggested how early connectivity development followed by response stabilization could generate the observed trajectory. These results link belief updating to integrated information in living neuronal networks and provide an empirical point of contact between IIT and the FEP.}


\maketitle

\section{Introduction}\label{sec:introduction}

Contemporary debates on the nature of consciousness have been shaped by two influential frameworks: Integrated Information Theory (IIT) \cite{tononi2004an, tononi2008consciousness, balduzzi2008integrated, oizumi2014from, albantakis2023integrated} and the Free-Energy Principle (FEP) \cite{friston2010the}. 
IIT, grounded in phenomenology, holds that consciousness is identical to a system’s integrated causal structure---an irreducible cause--effect repertoire quantified by $\Phi$---which specifies how experience exists here and now as an intrinsic property of the system.
In contrast, the FEP provides a normative account of self-organizing living systems, proposing that agents must minimize variational free energy (VFE) to constrain sensory surprise. 
This framework unifies perception, learning, and action under variational Bayesian inference and active inference. 
Within this view, deep generative models that enable long-horizon prediction confer adaptive advantages, suggesting why informational structures associated with consciousness may emerge.

Taken together, these perspectives suggest a complementary path toward synthesis.
IIT provides a proximate explanation, identifying conscious experience with integrated information structure itself, whereas FEP-based theories of consciousness (e.g., \cite{solms2018how, solms2019hard, rudrauf2017mathematical, williford2018the, whyte2021predictive}) offer an ultimate explanation in terms of teleology and adaptive function, echoing Tinbergen’s classic distinction between the proximate and ultimate causes \cite{tinbergen1963on}.
Proposals of conceptual bridges between the two frameworks are relatively recent.
For example, Markovian monism highlights formal parallels between IIT’s complexes and FEP’s Markov-blanketed agents, both of which insulate internal dynamics while mediating perception--action exchanges \cite{friston2020sentience}. 
Similarly, Integrated World Modeling Theory (IWMT) further argues that richly unified internal models---those with higher $\Phi$---are favored under active inference because they support long-term free-energy minimization \cite{safron2020an}. 
Consistent with this view, simulation studies have reported that evolving agents exhibit decreasing surprise alongside increasing $\Phi$ \cite{lundbakolesen2023phi}.
Collectively, these lines of research motivate a unified account in which integrated informational structure serves simultaneously as the substrate of experience (IIT) and as an emergent outcome of adaptive inferential dynamics (FEP).

Nevertheless, several important gaps remain. 
First, most evidence for an IIT--FEP association derives from theoretical or simulation studies: direct neural evidence from living systems is still scarce.
Second, the often-postulated negative correlation between $\Phi$ and VFE lacks mechanistic grounding and may not consistently hold, as the moment-to-moment relationships between $\Phi$ and surprise can vary in sign within a single task \cite{lundbakolesen2023phi}.
Finally, it remains unsolved how intrinsically integrated information both arises and operates during variational Bayesian inference under the FEP. 

In this study, we address these gaps by employing \textsl{in vitro} dissociated neuronal cultures grown on high-density multielectrode arrays (HD-MEAs).
Previous works have shown that such cultures, when driven by structured inputs, perform perceptual inference consistent with the FEP and can be modeled by canonical neural networks whose cost function is asymptotically equivalent to VFE \cite{isomura2015cultured,  isomura2020reverseengineering, isomura2022canonical, isomura2023experimental}. 
Building on this framework, we repeatedly presented stimuli generated by hidden signal sources and recorded spiking activity across successive sessions.
From these data, we estimated VFE and its decomposition into Bayesian surprise (complexity) and accuracy.
To obtain proxy measures of integrated information, we computed pairwise synergistic information ($\Phi_R$) \cite{mediano2019beyond} and constructed weighted graphs, from which main complexes were extracted using minimum-cut procedures inspired by IIT 2.0-style analyses \cite{kitazono2020efficient, kitazono2023bidirectionally}.

Based on these considerations, we address the following questions.
First, does integrated information necessarily accompany a decrease in VFE, or does it instead track other FEP-related quantities? 
Second, how does integrated information evolve as networks improve inference---does it increase monotonically, remain stable, or follow a non-linear trajectory?
Finally, if a consistent evolution pattern is observed, how can it be functionally interpreted?
Our aim is to answer these questions and to advance the IIT--FEP dialogue from theoretical plausibility to empirical grounding by jointly quantifying FEP-related quantities and integrated informational structure in living neural networks.
Rather than directly assessing subjective experience, we ask how a learning process in a living neuronal network appears through the complementary lenses of IIT and the FEP.
Accordingly, our claims are limited to the behavior of IIT-inspired proxies during FEP-consistent learning, rather than to full IIT quantification or direct evidence for consciousness.
In doing so, we seek to provide empirical constraints for future attempts to link IIT's proximate account of integrated information with the FEP's functional account of adaptive inference, and thereby to inform broader frameworks for the mechanisms and adaptive roles of consciousness.
In addition to addressing these empirical questions, we developed a mechanistic analysis based on an Ising neural Bayesian model to interpret the observed coupling between belief updating and integrated information, to account for the hill-shaped trajectory of \(\Phi\), and to identify sufficient conditions under which these phenomena can arise.
\section{Results}

\begin{figure}[t!]
  \centering
  \includegraphics[
  draft=false,
  width=1.0\linewidth,
  height=0.45\textheight,
  keepaspectratio
  ]{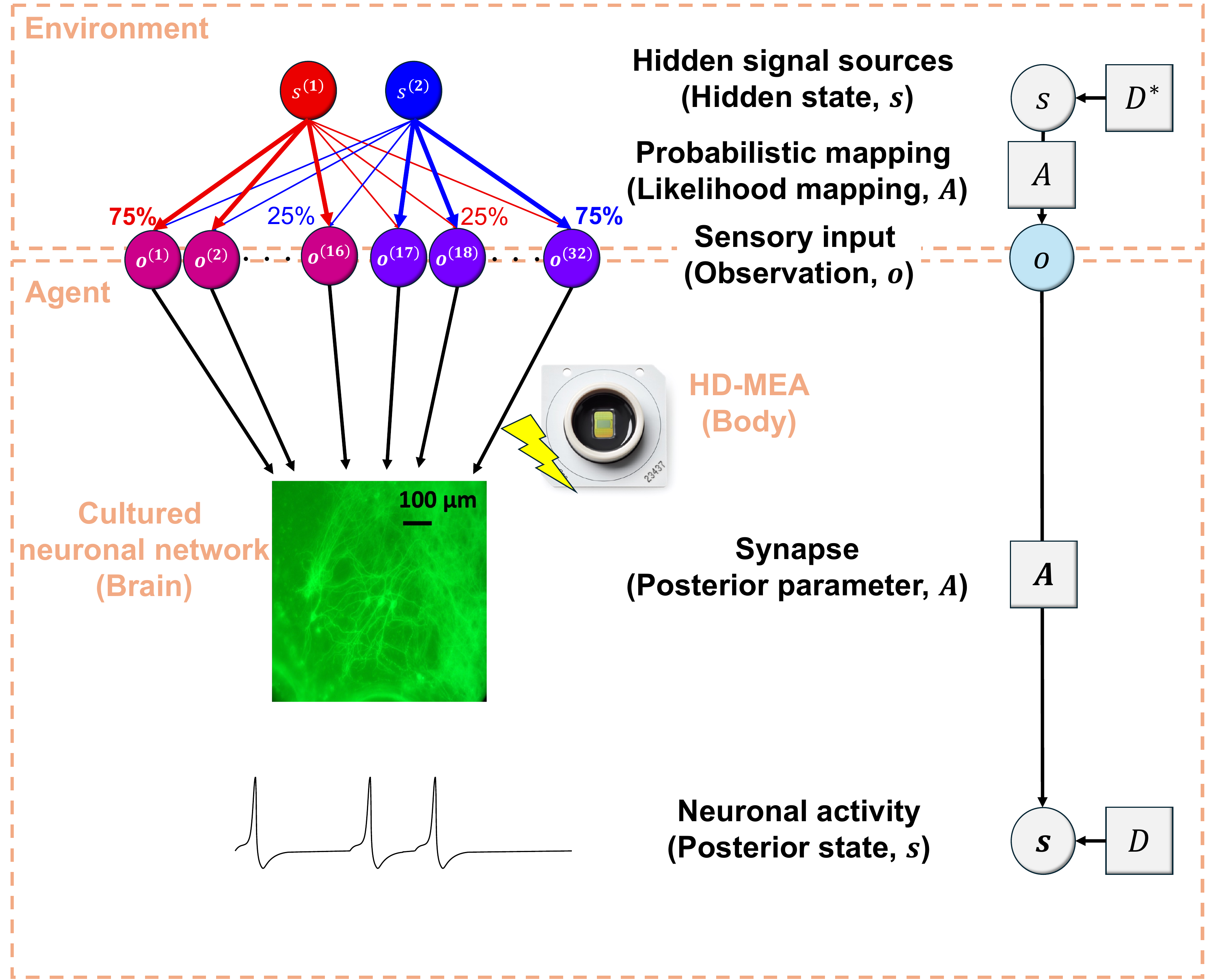}
  \captionsetup{
  font={small,stretch=1.0},
  labelfont=bf,
  justification=raggedright,
  singlelinecheck=false,
  skip=4pt
  }
  \caption{\textbf{Experimental paradigm.} Setup (left) and corresponding POMDP (right), following the design of prior research \cite{isomura2023experimental}.
  In each trial, hidden signal sources $s = (s^{(1)}, s^{(2)})$ in a computer stochastically generate observations $o$ through a likelihood mapping $A$.
  These hidden sources were not directly observable to the cultured neuronal network, whereas the observations delivered via 32 electrodes on the HD-MEA were directly observable.
  These electrical stimuli evoked synaptically mediated responses, corresponding to posterior states $\boldsymbol{s}$ mediated by the posterior parameter $\boldsymbol{A}$.
  }
  \label{fig:fig1}
\end{figure}

\subsection{Study aims and experimental paradigm}

To bridge IIT and the FEP within a living neural system, we examined how integrated information emerges and functions within a form of self-organization suggested to follow the FEP.
We employed dissociated neuronal cultures grown on HD-MEAs, using a repeated-stimulation paradigm in which probabilistic observations generated by two hidden signal sources were delivered via 32 electrodes (Fig.\ref{fig:fig1}, left).
Previous studies have shown that such cultures acquire the capacity to infer hidden sources, with VFE---empirically computed from a canonical neural network formulation---decreasing during learning \cite{isomura2015cultured, isomura2020reverseengineering, isomura2022canonical, isomura2023experimental}. 
Building on this design, we recorded spiking activity as networks inferred and learned, computed FEP-related quantities (VFE, Bayesian surprise, and accuracy), and derived proxy measures of integrated information ($\Phi_R^{\text{mc}}$ and coreness) to analyze their trajectories and interrelationships during perceptual inference. 

Within the FEP framework, VFE in variational Bayesian inference under a generative process modeled as a partially observable Markov decision process (POMDP; Fig.\ref{fig:fig1}, right) can be written as follows: 
\begin{equation}
    F(Q(s, A), o) = \underbrace{D_{\textup{KL}} (Q(s, A) \parallel P(s, A))}_{\text{complexity (Bayesian surprise) }} - \underbrace{\mathbb{E}_{Q(s, A)}[\ln{P(o \mid s, A)}]}_{\text{Accuracy}}.
\end{equation}
The canonical neural network \cite{isomura2020reverseengineering, isomura2022canonical} is mathematically equivalent to variational Bayes in this setting, enabling the empirical estimates of VFE, Bayesian surprise, and accuracy directly from the recorded activity.

The generative process comprised two independent binary signal sources $s^{(1)}$ and $s^{(2)}$, which stochastically generated 32 binary observations through a 0.75/0.25 likelihood mapping across channel halves.
Each observation was delivered to the culture as an electrical pulse (Fig.\ref{fig:fig1}, left).
One experiment consisted of 100 sessions, each comprising 256 trials presented at 1-s intervals, with a 244-s rest period between sessions (see \hyperref[method:electrophysiology]{Methods 'Electrophysiological experiments' section}, for details).

\begin{figure}[t!]
  \centering
  \includegraphics[
  draft=false,
  width=1.0\linewidth,
  height=0.45\textheight,
  keepaspectratio
  ]{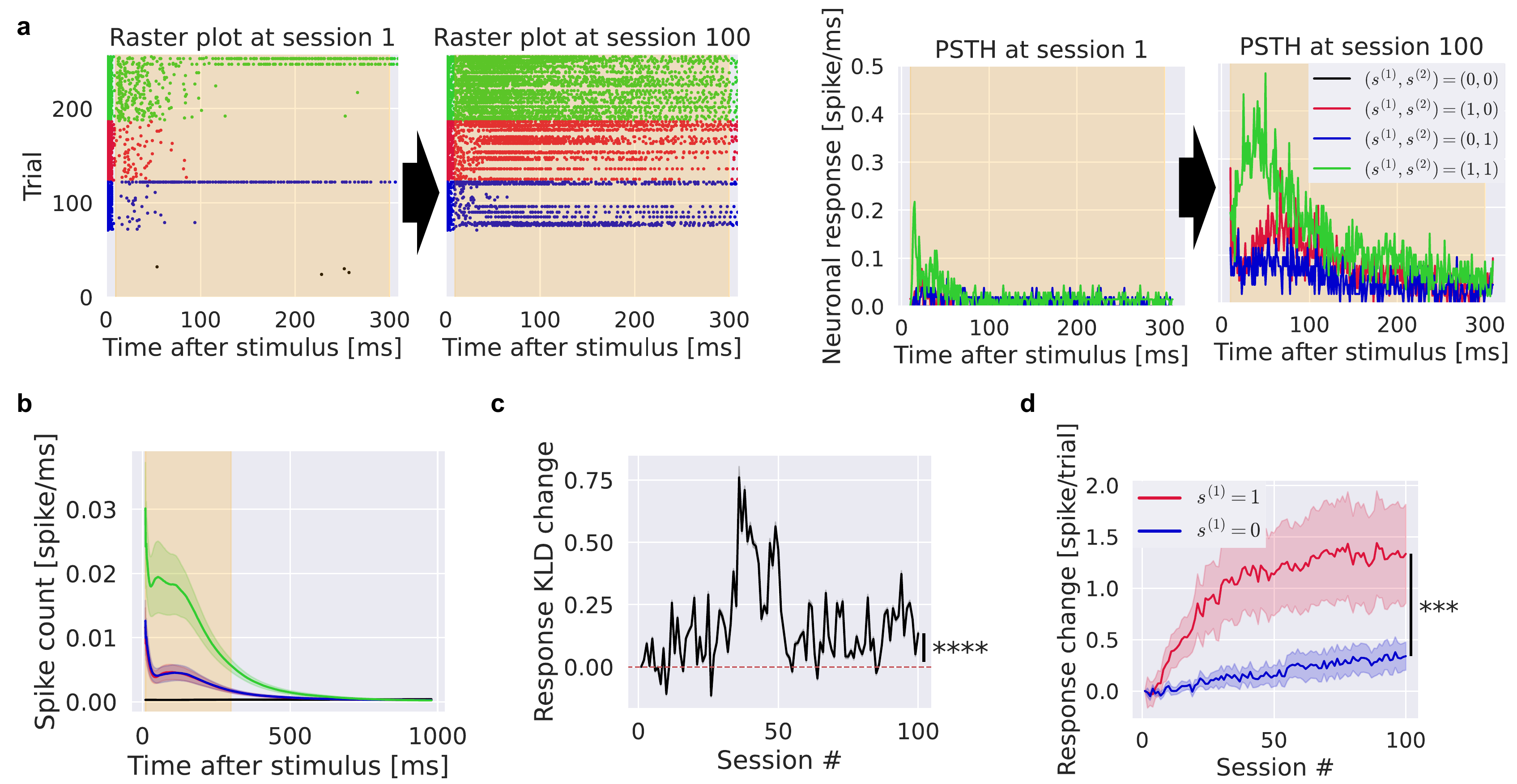}
  \captionsetup{
  font={small,stretch=1.0},
  labelfont=bf,
  justification=raggedright,
  singlelinecheck=false,
  skip=4pt
  }
  \caption{\textbf{Perceptual inference by neuronal networks.} (a) Changes in neuronal activity at a single representative electrode across sessions.
  Colors indicate hidden source states.
  (Left) Raster plots of spiking activity across 256 trials in the first and last sessions.
  The horizontal axis denotes time after electrical stimulation (ms), and the vertical axis denotes trials, sorted by hidden source states.
  Each dot represents a spike detected at the electrode.
  (Right) Post-stimulus time histograms (PSTHs) from the first and last sessions.
  The horizontal axis denotes time after stimulation and the vertical axis shows the mean spike counts.
  (b) PSTH averaged across sessions, electrodes, and experiments. 
  A peak is evident at $\sim$100--200 ms post-stimulation.
  (c) Change from the first session in the Kullback--Leibler divergence (KLD) between the distributions of evoked spike counts for trials with $(s^{(1)},s^{(2)})=(1,0)$ and $(s^{(1)},s^{(2)})=(0,1)$, averaged across electrodes.
  KLD significantly increased in the final session (Wilcoxon signed-rank test; final session, $n=7{,}613$ electrodes from 27 experiments, $\text{****}p=2.7\times10^{-144}< 0.001$).
  (d) Change from the first session in the mean evoked spike count of $s^{(1)}$-preferring electrodes when $s^{(1)}=1$ versus $s^{(1)}=0$, averaged across experiments.
  Responses when $s^{(1)}=1$ grew significantly more than those when $s^{(1)}=0$ (Wilcoxon signed-rank test; final session, $n=27$, $\text{***}p=2.5\times10^{-3} < 0.005$).
  }
  \label{fig:fig2}
  \phantomsubcaption\label{fig2:A}
  \phantomsubcaption\label{fig2:B}
  \phantomsubcaption\label{fig2:C}
  \phantomsubcaption\label{fig2:D}
  
\end{figure}
\begin{figure}[t!]
  \centering
  \includegraphics[
  draft=false,
  width=0.95\linewidth,
  height=0.45\textheight,
  keepaspectratio
  ]{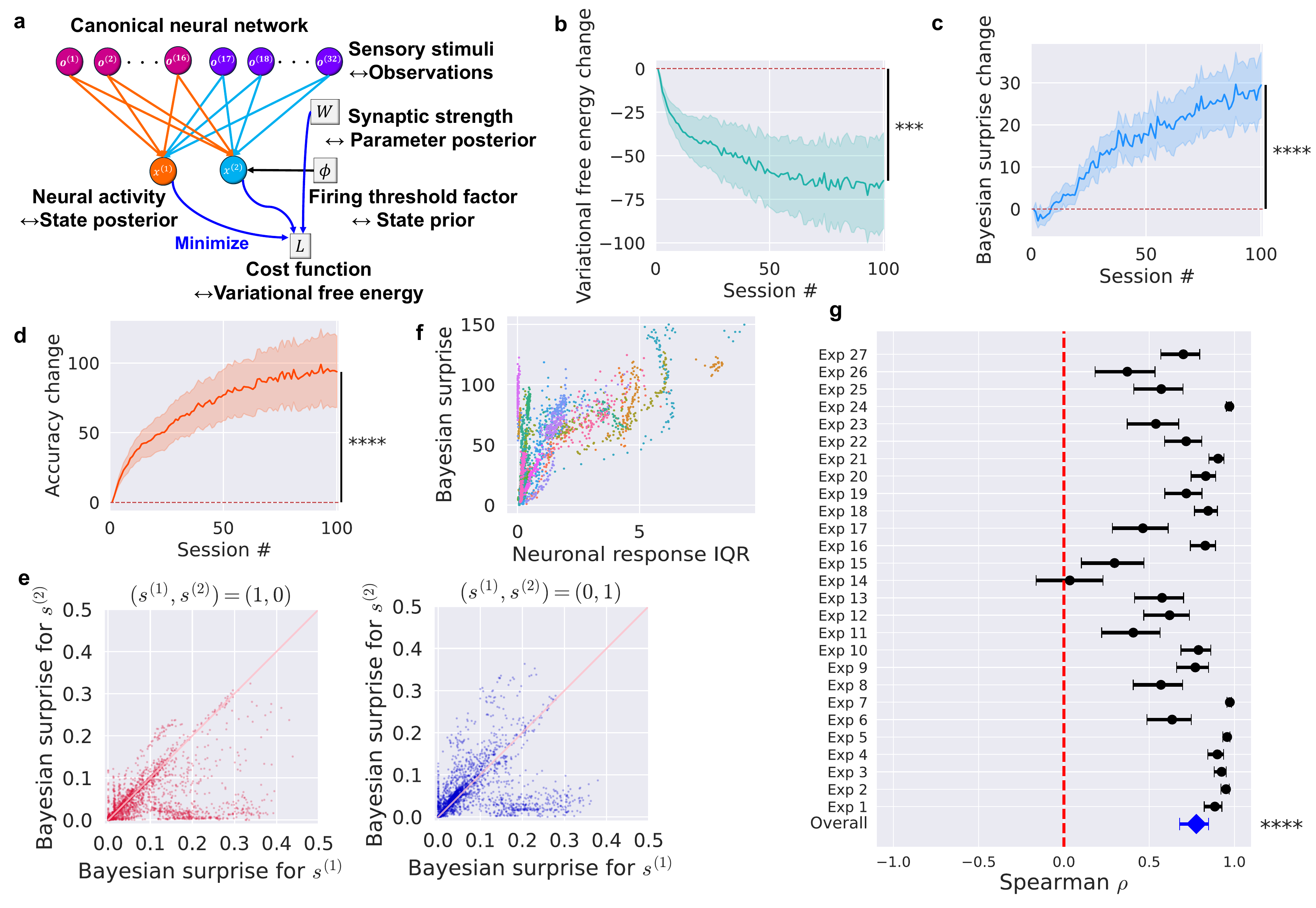}
  \captionsetup{
  font={small,stretch=1.0},
  labelfont=bf,
  justification=raggedright,
  singlelinecheck=false,
  skip=4pt
  }
  \caption{\textbf{Variational Bayes formulation.} (a) Schematic of a canonical neural network.
  Neural activity $x$ is determined by sensory input $o$ through synaptic weights $W$ and a firing threshold factor $\phi$.
  Assuming that the dynamics of $x$ and $W$ minimize a common cost function $L$, the network is mathematically equivalent to variational Bayesian inference in the POMDP framework shown in Fig.\ref{fig:fig1}.
  Specifically, sensory inputs correspond to observations $o$, synaptic strengths to parameter posteriors $\boldsymbol{A}$, threshold factors to state priors $D$, and neural activity to state posteriors $\boldsymbol{s}$.
  (b--d) Changes from the first session in VFE, Bayesian surprise, and accuracy, respectively, averaged across experiments.
  VFE significantly decreased, whereas Bayesian surprise and accuracy significantly increased (Wilcoxon signed-rank test; final session, $n=27$, $\text{***}p = 1.4\times10^{-3}<0.005$, $\text{****}p=6.0\times10^{-4}<0.001$, and $\text{****}p=1.5\times10^{-8}<0.001$, respectively).
  (e) Distributions of mean $s^{(1)}$ Bayesian surprise and mean $s^{(2)}$ Bayesian surprise within a session for trials with $(s^{(1)},s^{(2)})=(1,0)$ (left) and $(0, 1)$ (right).
  Each point represents one session, yielding 2,700 points across 27 experiments. 
  For $(1,0)$, the $s^{(1)}$ Bayesian surprise was significantly greater than the $s^{(2)}$ Bayesian surprise (two-sided binomial test on the sign of paired differences; $k=1{,}712, n=2{,}700, p=1.6\times 10^{-44}$).
  Conversely, for $(0,1)$, the $s^{(2)}$ Bayesian surprise was significantly greater ($k=1{,}487, n=2{,}700, p=1.5\times 10^{-7}$).
  (f) Scatter plot of the interquartile range (IQR) of mean evoked responses of preferring electrodes versus Bayesian surprise.
  Each point represents one session ($2,700$ points in total) with colors indicating different experiments.
  (g) Spearman correlation coefficients between neuronal response IQR and Bayesian surprise with 95\% confidence intervals for each experiment, and their meta-analysis using the DerSimonian--Laird method.
  Shown are the Fisher-$z$-transformed mean correlation under the random-effects model, its 95\% confidence interval, and the 95\% prediction interval.
  The mean correlation was significantly positive (two-sided $Z$-test; $\text{****}p = 4.4\times10^{-22} < 0.001$).
  }
  \label{fig:fig3}
  \phantomsubcaption\label{fig3:A}
  \phantomsubcaption\label{fig3:B}
  \phantomsubcaption\label{fig3:C}
  \phantomsubcaption\label{fig3:D}
  \phantomsubcaption\label{fig3:E}
  \phantomsubcaption\label{fig3:F}
  \phantomsubcaption\label{fig3:G}
  
\end{figure}

\subsection{Perceptual inference by neuronal networks}

We conducted 27 experiments across 12 independently prepared cultures using an HD-MEA (26,400 electrodes; up to 1,024 recorded simultaneously at a sampling rate of 20 kHz; 32 stimulation channels and $\leq$992 recording channels). 
Spike rasters and PSTHs at a representative electrode revealed source-selective responses that strengthened progressively from session 1 to session 100 (Fig.\ref{fig2:A}).
Across electrodes and experiments, spike counts peaked around 100--200 ms post-stimulation; thus, the number of spikes within a 10--300 ms window was defined as the evoked response (Fig.\ref{fig2:B}).
Across electrodes, the Kullback-Leibler divergence (KLD) of the responses between $(s^{(1)},s^{(2)})=(1,0)$ and $(s^{(1)},s^{(2)})=(0, 1)$ increased significantly (Fig.\ref{fig2:C}).
Moreover, when tracking the changes in the average evoked responses of $s^{(1)}$-preferring electrodes (those selectively responsive to $s^{(1)}$), responses grew more strongly during trials in which $s^{(1)}$ was active, demonstrating the emergence and reinforcement of source selectivity under repeated, source-generated stimulation (Fig.\ref{fig2:D}).
Together, these findings indicate source-selective encoding consistent with perceptual inference by neuronal networks: the cultures became sensitive to the hidden signal sources’ states despite receiving probabilistically generated electrical stimulation.

\subsection{Canonical neural network and variational Bayesian inference}

We formalized the inference using a canonical neural network \cite{isomura2020reverseengineering, isomura2022canonical} (Fig.\ref{fig3:A}), which is mathematically equivalent to variational Bayesian inference under the POMDP (Fig.\ref{fig:fig1}, right).
This formulation allowed empirical evaluation of VFE, its complexity term (Bayesian surprise), and accuracy from recorded neuronal responses and inferred parameters (see \hyperref[method:FEP]{Methods 'FEP-based analysis' section}, for details).
Across experiments, VFE decreased, whereas both Bayesian surprise and accuracy increased significantly (Fig.\ref{fig3:B}--\ref{fig3:D}), consistent with self-organization under the FEP and reflecting enhanced belief updating and model complexity. 
Since accuracy increased more strongly than Bayesian surprise, VFE—which is the latter minus the former—decreased across sessions.
We further decomposed Bayesian surprise by source and found it to be selectively larger for the currently true source (two-sided binomial sign tests; Fig.\ref{fig3:E}).
Moreover, Bayesian surprise was strongly coupled to response diversity quantified by the session-wise interquartile range (IQR) of evoked activity (mean $\rho = 0.777$, 95\% CI [0.679, 0.848], $\tau^2=0.298$, $Q=626.7$, $I^2 = 95.9$, and $p=4.4\times10^{-22}$; meta-analysis on Spearman correlations under a random-effects model; Fig.\ref{fig3:F}, \ref{fig3:G}).

\begin{figure}[t!]
  \centering
  \includegraphics[
  draft=false,
  width=1.0\linewidth,
  height=0.45\textheight,
  keepaspectratio
  ]{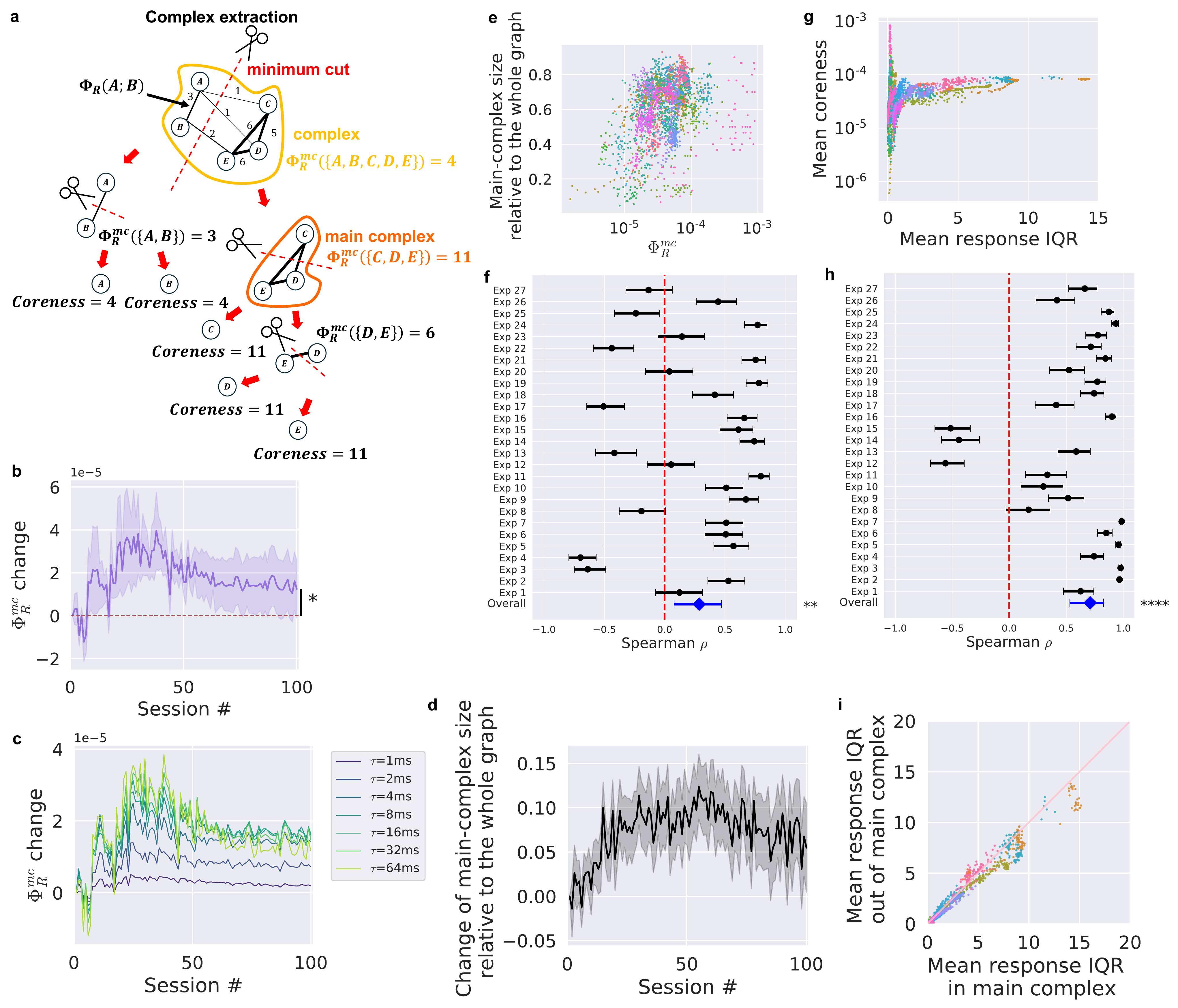}
  \captionsetup{
  font={small,stretch=1.0},
  labelfont=bf,
  justification=raggedright,
  singlelinecheck=false,
  skip=4pt
  }
  \caption{\textbf{Integrated information.} (a) Weighted undirected graphs were constructed by computing $\Phi_R$ between all pairs of preferring electrodes, yielding one graph per session. 
  Each graph was recursively partitioned using minimum cuts until single vertices remained. 
  For each vertex set, the sum of edge weights crossing the minimum cut was defined as $\Phi_R^{\text{mc}}$. 
  Based on $\Phi_R^{\text{mc}}$, complexes and main complexes were identified, and a coreness value was assigned to each vertex.
  (b) Change from the first session in $\Phi_R^{\text{mc}}$ maximized across the temporal coarse-graining scales $\tau$ (time window widths for transition-probability estimation) for each session, averaged across experiments.
  The final-session increase was significant (Wilcoxon signed-rank test; final session, $n=27$, $\text{*}p=4.1\times10^{-2}<0.05$).
  (c) Change from the first session in $\Phi_R^{\text{mc}}$ for each temporal scale $\tau$, averaged across experiments. 
  (d) Change from the first session in the ratio of the number of vertices in the main complex to the total number of vertices.
  (e) Scatter plot of main-complex $\Phi_R^{\text{mc}}$ versus the ratio of vertices in the main complex.
  (f) Spearman correlations between main-complex $\Phi_R^{\text{mc}}$ and the ratio of vertices in the main complex, with 95\% confidence intervals and meta-analysis across experiments.
  A significant positive correlation was observed ($\text{**}p = 7.8\times10^{-3}<0.01$).
  (g) Scatter plot of the mean IQR of neuronal responses across all electrodes versus the mean coreness across all electrodes.
  (h) Spearman correlations between mean neuronal response IQR and mean coreness across electrodes, with 95\% confidence intervals and meta-analysis.
  A significant positive correlation was observed ($\text{****}p = 3.4\times10^{-9}<0.001$).
  (i) Scatter plot comparing the mean neuronal response IQR of electrodes inside versus outside the main complex.
  The pink line indicates the identity line.
  The mean IQR inside the main complex was significantly larger (Wilcoxon signed-rank test; $n=2{,}700$, $\text{****}p=4.0\times10^{-70}<0.001$).
  }
  \label{fig:fig4}
  \phantomsubcaption\label{fig4:A}
  \phantomsubcaption\label{fig4:B}
  \phantomsubcaption\label{fig4:C}
  \phantomsubcaption\label{fig4:D}
  \phantomsubcaption\label{fig4:E}
  \phantomsubcaption\label{fig4:F}
  \phantomsubcaption\label{fig4:G}
  \phantomsubcaption\label{fig4:H}
  \phantomsubcaption\label{fig4:I}
\end{figure}

\subsection{Integrated information and informational cores within neuronal networks}

To track integrated information during learning, we constructed weighted graphs for each session by computing $\Phi_R$ \cite{mediano2019beyond}---an empirical measure of synergistic information \cite{varley2023decomposing, luppi2024a}---between all pairs of preferring electrodes, and then extracted complexes using a minimum-cut procedure \cite{kitazono2020efficient}.
Because $\Phi_R$ is estimated from state transition probabilities, it depends on the temporal coarse-graining scale. 
Here, $\tau$ denotes the width of the time window used to coarse-grain post-stimulus activity into binary states for transition-probability estimation; we evaluated $\tau=1,2,4,8,16,32$, and $64$ ms.
State transition probabilities for $\Phi_R$ estimation were derived exclusively from stimulation trials, following the perturbational approach recommended by IIT, to better capture cause--effect power elicited by exogenous inputs. 
For each subgraph, $\Phi_R^{\text{mc}}$ was defined as the sum of edge weights crossing the minimum cut.
By comparing these values with those of its subsets or supersets, each subgraph was classified as a complex, main complex, or neither (see \hyperref[method:complex_extraction]{Methods 'Complex extraction' section}, for details).
The $\Phi_R^{\text{mc}}$ of the main complex indexed integrated information, while coreness \cite{kitazono2023bidirectionally} quantified each node's contribution to informational cores (Fig.\ref{fig4:A}).
For each session, we selected the value of $\tau$ that maximized main-complex $\Phi_R^{\text{mc}}$ and used the corresponding $\Phi_R^{\text{mc}}$ as the primary session-wise measure.

Using, for each session, the temporal scale $\tau$ that maximized $\Phi_R^{\text{mc}}$, $\Phi_R^{\text{mc}}$ exhibited a hill-shaped trajectory, rising early and then declining or stabilizing at a lower level (Fig.\ref{fig4:B}).
Qualitatively similar rise-and-fall patterns were also observed when each temporal scale was analyzed separately (Fig.\ref{fig4:C}), suggesting that the trajectory was not solely driven by the session-wise selection of $\tau$.
To formally assess the nonlinear temporal structure of $\Phi_R^{\text{mc}}$, we performed exploratory within-experiment regression analyses.
The dependent variable was the z-scored session-wise $\Phi_R^{\mathrm{mc}}$, and the predictor was z-scored session number $t$ and its square $t^2$.
Random-effects meta-analysis showed that the regression coefficient of $t^2$ was significantly negative (Supplementary Fig.~S4), supporting the non-monotonic concave-down trajectory.
Main-complex size followed a similar expansion--contraction profile (Fig.\ref{fig4:D}).
$\Phi_R^{\text{mc}}$ scaled positively with main-complex size (mean $\rho=0.286$, 95\% CI [0.077, 0.471], $Q=734.2$, $I^2 = 96.5$, $\tau^2=0.318$, and $p=7.8\times10^{-3}$; Fig.\ref{fig4:E}, \ref{fig4:F}). 
Response diversity also increased with mean coreness (mean $\rho=0.709$, 95\% CI [0.531, 0.827], $Q=1239.2$, $I^2 = 97.9$, $\tau^2=0.594$, and $p=3.4\times10^{-9}$) and was significantly higher inside than outside the main complex (Wilcoxon signed-rank test; $n = 2{,}700$ session pairs, $p=4.0\times10^{-70}$) (Fig.\ref{fig4:G}--\ref{fig4:I}). 
Together, these results suggest that higher integrated information is accompanied by larger informational cores that concentrate diverse neuronal activity. 

\begin{figure}[t!]
  \centering
  \includegraphics[
  draft=false,
  width=1.0\linewidth,
  height=0.45\textheight,
  keepaspectratio
  ]{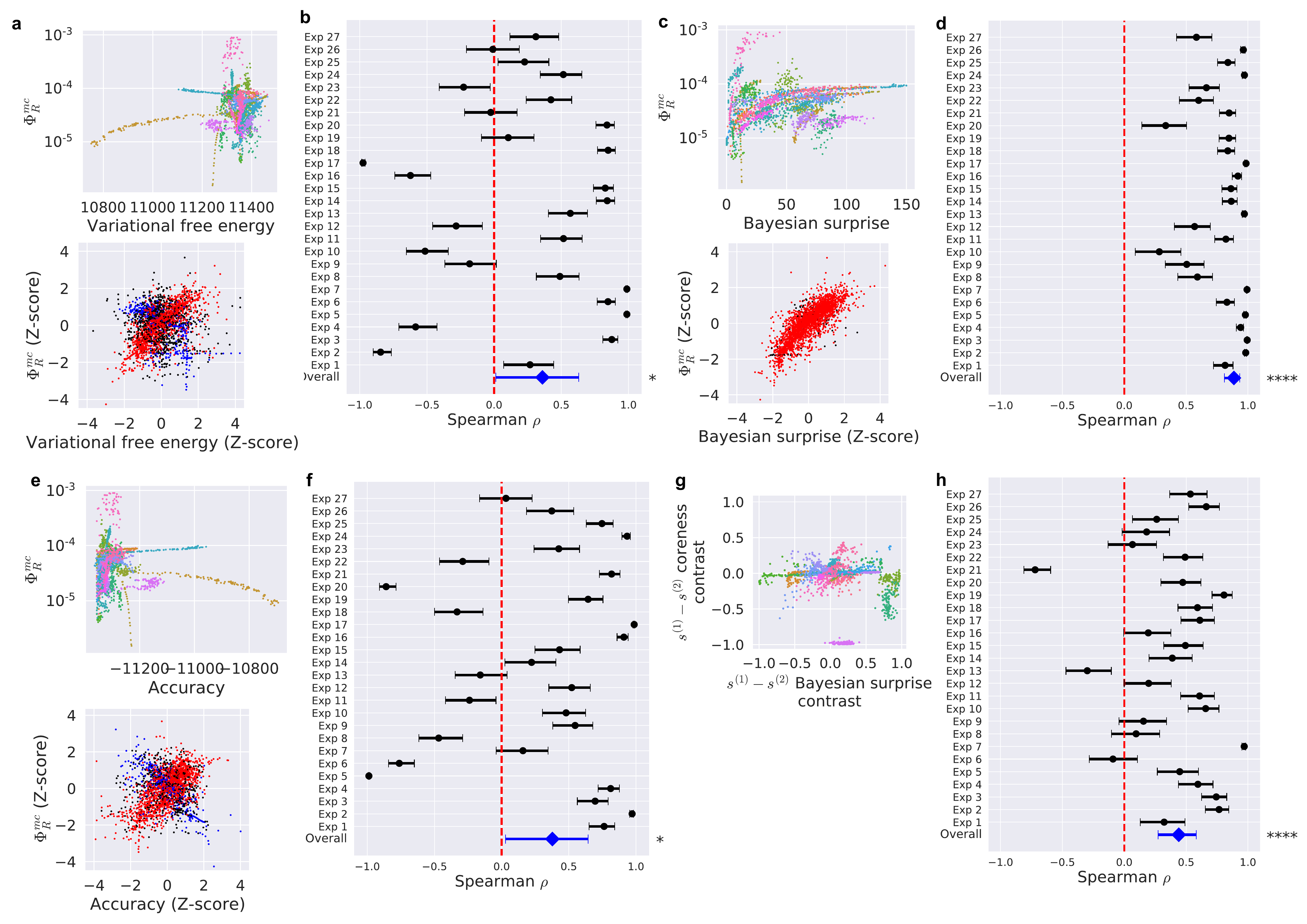}
  \captionsetup{
  font={small,stretch=1.0},
  labelfont=bf,
  justification=raggedright,
  singlelinecheck=false,
  skip=4pt
  }
  \caption{\textbf{Integrated information in perceptual inference.} (a) Scatter plot of VFE versus $\Phi_R^{\text{mc}}$.
  The upper panel shows raw values, with each point representing one session ($2{,}700$ points in total across experiments) and colors indicating different experiments. 
  The lower panel shows Z-scores; for each experiment, sessions were plotted in red if the Spearman correlation exceeded 0.3, in blue if less than $-$0.3, and in black otherwise.
  (b) Spearman correlations between VFE and $\Phi_R^{\text{mc}}$ for each experiment with 95\% confidence intervals, and their meta-analysis. 
  A significant positive overall correlation was observed (two-sided $Z$-test; $\text{*}p = 4.2\times10^{-2}<0.05$).
  (c) Scatter plot of Bayesian surprise versus $\Phi_R^{\text{mc}}$, in the same format as (a).
  (d) Spearman correlations between Bayesian surprise and $\Phi_R^{\text{mc}}$ for each experiment with 95\% confidence intervals, and their meta-analysis. 
  A significant positive correlation was observed (two-sided $Z$-test; $\text{****}p=5.0\times10^{-23} < 0.001$).
  (e) Scatter plot of accuracy versus $\Phi_R^{\text{mc}}$, in the same format as (a).
  (f) Spearman correlations between accuracy and $\Phi_R^{\text{mc}}$ for each experiment with 95\% confidence intervals, and their meta-analysis. 
  A significant positive correlation was observed (two-sided $Z$-test; $\text{*}p=3.4\times10^{-2} < 0.05$).
  (g) Scatter plot of the contrast between the mean coreness of $s^{(1)}$-preferring versus $s^{(2)}$-preferring electrodes against the contrast between $s^{(1)}$ Bayesian surprise and $s^{(2)}$ Bayesian surprise.
  (h) Spearman correlations between coreness contrast and Bayesian surprise contrast for each experiment with 95\% confidence intervals, and their meta-analysis. 
  A significant positive correlation was observed (two-sided $Z$-test; $\text{****}p=5.9\times10^{-9} < 0.001$).
  }
  \label{fig:fig5}
  \phantomsubcaption\label{fig5:A}
  \phantomsubcaption\label{fig5:B}
  \phantomsubcaption\label{fig5:C}
  \phantomsubcaption\label{fig5:D}
  \phantomsubcaption\label{fig5:E}
  \phantomsubcaption\label{fig5:F}
  \phantomsubcaption\label{fig5:G}
  \phantomsubcaption\label{fig5:H}
\end{figure}

\subsection{Integrated information during perceptual inference under the FEP}

We next examined the relationship between $\Phi_R^{\text{mc}}$ and FEP-related quantities across all sessions. 
Bayesian surprise showed the most robust positive association with $\Phi_R^{\text{mc}}$ across experiments, whereas accuracy and VFE showed more modest and heterogeneous positive associations ($\Phi_R^{\text{mc}}$--VFE: mean $\rho=0.360$, 95\% CI [0.013, 0.629], $Q=1996.6$, $I^2 = 98.7$, $\tau^2=0.918$, and $p=4.2\times10^{-2}$; $\Phi_R^{\text{mc}}$--Bayesian surprise: mean $\rho=0.886$, 95\% CI [0.810, 0.933], $Q=1043.4$, $I^2 = 97.5$, $\tau^2=0.532$, and $p=5.0\times10^{-23}$; $\Phi_R^{\text{mc}}$--Accuracy: mean $\rho=0.378$, 95\% CI [0.029, 0.645], $Q=2027.4$, $I^2 = 98.7$, $\tau^2=0.942$, and $p=3.4\times10^{-2}$; Fig.\ref{fig5:A}--\ref{fig5:F}).
Thus, integrated information was most consistently associated with belief updating, whereas its positive relationships with accuracy and VFE were weaker and more heterogeneous.

To further characterize the dependency structure among $\Phi_R^{\text{mc}}$, Bayesian surprise, and accuracy, we performed exploratory within-experiment regression analyses using z-scored session-wise variables (Supplementary Fig.~S1 and Supplementary Table~S1).
Bayesian surprise showed more consistent positive coefficients than accuracy in both univariate and multivariable models, and models containing Bayesian surprise generally accounted for more session-wise variation in $\Phi_R^{\text{mc}}$ than the corresponding accuracy-only models (Supplementary Fig.~S1 and Supplementary Table~S1).
Random-effects meta-analyses of the regression coefficients further showed positive Bayesian-surprise coefficients across Bayesian-surprise-containing models (Supplementary Fig.~S2), whereas accuracy coefficients were smaller and less consistently positive (Supplementary Fig.~S3).
This pattern was also observed after including session number as an additional covariate, suggesting that the association between Bayesian surprise and $\Phi_R^{\text{mc}}$ was not solely attributable to monotonic session progression.

Finally, the contrast in coreness between $s^{(1)}$- and $s^{(2)}$-preferring electrodes tracked the contrast in source-specific Bayesian surprise (mean $\rho = 0.441$, 95\% CI [0.275, 0.581], $Q=611.3$, $I^2 = 95.7$, $\tau^2=0.268$, and $p=5.9\times10^{-9}$; Fig.\ref{fig5:G}, \ref{fig5:H}). 
In other words, stronger belief updating was associated with greater contributions to informational cores, linking the content of inference to the geography of integration within the same network. 

\subsection{Theoretical interpretation: shared critical modes}
To understand why Bayesian surprise and integrated information covary, and integrated information follows a hill-shaped trajectory in our experiments, we constructed a theoretical model based on an Ising neural Bayesian model (see Supplementary Note for the derivations and additional theoretical details).
Note that while we use the model to derive expressions for Bayesian surprise and integrated information, these are not identical to the complexity term of the canonical neural network or \(\Phi_R^{\mathrm{mc}}\), which are computed from actual neural activity.
Instead, they are introduced to capture generic belief updating and partition-sensitive integration in a form that is analytically tractable within the model.
Also note that this is an independent mathematical model; we did not fit this model to the present data.
Rather, its aim is to provide a possible mechanism underlying our results.

In this model, neuronal activity is assumed to represent the posterior beliefs over binary hidden states \(s\in\{-1,1\}^N\), and the posterior distribution is represented by an Ising distribution:
\[
P_o(s)\propto\exp\left[h(o)^\top s + \frac12 s^\top K s\right],
\]
where \(h(o)\) is an observation-dependent local field and \(K\) is a symmetric interaction matrix.

The Bayesian surprise is the KL divergence from the posterior belief to the prior belief $P_p$:
\[
B(o) = D_{\mathrm{KL}}(P_o \parallel  P_p) = \sum_{s\in \{-1, 1\}^N} P_o(s) \log \frac{P_o(s)}{P_p(s)}.
\]
For a small observation-induced perturbation \(\delta h(o) = h(o) - h_p\), we have the approximated expression
\[
B(o) \sim \frac12 \delta h(o)^\top F(h(o)) \delta h(o),
\]
where $F(h(o))$ is the Fisher information with respect to $h(o)$. 
Letting $m(o)$ denote the mean state, the Fisher information is also represented by the Jacobian matrix $\partial m / \partial h^\top$.

Because the exact posterior mean \(m(o)=E_{P_o}[S]\) is generally difficult to compute, we use a naive mean-field approximation. 
This approximation replaces the random coupling input to each unit by its posterior mean, yielding
\[
m^\ast(o)=\tanh\{h(o)+Km^\ast(o)\}.
\]
Thus the mean-field fixed point \(m^\ast(o)\) is used as an approximation to \(m(o)\), rather than as an exact identity.
Given that $D(o) = \mathrm{diag}\left(\sqrt{1-m_1^*(o)^2}, ..., \sqrt{1-m_N^*(o)^2}\right)$ and $M(o)=D(o)KD(o)$, the local mean-field susceptibility $\partial m^* / \partial h^\top$ is expressed by
\[
\frac{\partial m^*}{\partial h^\top} = D(o)(I - M(o))^{-1}D(o).
\]
Thus, approximating $F(h(o))$ by $\partial m^* / \partial h^\top$ and using the eigenvalues $\lambda_k$ and the eigenvectors $q_k$ of $M(o)$, we have
\[
B(o)\simeq \frac12\sum_k \frac{|q_k^\top D(o)\delta h|^2}{1-\lambda_k}.
\]

In parallel, a linear--Gaussian integrated-information proxy $\Phi_\pi(o)$ for a partition $\pi$---defined as the Kullback--Leibler divergence between the full dynamics and dynamics with cross-partition couplings removed---has the approximation 
\[
\Phi_\pi(o)\simeq \frac12\sum_k \frac{\|\Delta_\pi M(o) q_k\|^2}{1-\lambda_k^2}, 
\] where \(\Delta_\pi  M(o)=M(o)-M_\pi(o)\) is the difference between the intact matrix \(M(o)\) and the \(\pi\)-partitioned matrix \(M_\pi(o)\).

Here we adopt the atomic partition \(\pi_\mathrm{atom}=\{\{1\},\{2\},\ldots,\{N\}\}\) to give
\[
\Phi_{\mathrm{atom}}(o)
=
\frac12
\sum_k
\frac{\lambda_k(o)^2}{1-\lambda_k(o)^2}.
\]

Near a positive critical mode $\lambda_k\to1$, $1-\lambda_k^2\simeq2(1-\lambda_k)$.
Thus the expressions obtained above show that \(B(o)\) and \(\Phi_\mathrm{atom}(o)\) can both be amplified by shared effective criticality.
This result suggests their strong empirical correlation arises when a positive critical mode dominates the dynamics. 
Note that this does not demonstrate that the cultures were empirically critical.

The numerators differ: \(B(o)\) reflects how the input-driven update is aligned with an effective mode, while \(\Phi_\pi(o)\) reflects how much a mode is broken by a partition.
Therefore, depending on the input and partition, the quantities need not always covary. 
Nevertheless, our theoretical analysis predicts that in regimes of large belief updating along positive critical modes, both metrics increase together, consistent with our experimental observations.

The same effective-mode picture also provides a qualitative mechanism for the hill-shaped trajectory of \(\Phi\) during learning.
We suppose the interaction matrix \(K\) slowly changes with learning.
Then the derivative of the critical eigenvalue with respect to the learning stage \(\ell\) is given by:
\[
\frac{d\lambda_\ast}{d\ell}
=
(Dq_\ast)^\top
\frac{dK}{d\ell}
(Dq_\ast)
-
2\lambda_\ast
\sum_i
\frac{m_i^*\,dm_i^*/d\ell}{1-{m_i^*}^2}
q_{\ast i}^2
\]

Early in learning, responses are still weakly stabilized, so \(m_i^*\simeq0\), \(R_i=1-{m_i^*}^2\simeq1\). 
As the effective interaction connectivity \(K\) increases, the first term dominates the second term, and the dominant positive eigenvalue increases.
Later in learning, responses become more stabilized and selective.
In the binary idealization, \(m_i^*\) moves toward its selected state \(\pm1\), and by reducing \(R_i\) the second term dominates the first term, causing the eigenvalue to decrease.
This provides a possible mechanism by which integrated information can rise during connectivity development and fall after response stabilization.

\begin{figure}[t!]
  \centering
  \includegraphics[
  draft=false,
  width=1.0\linewidth,
  height=0.45\textheight,
  keepaspectratio
  ]{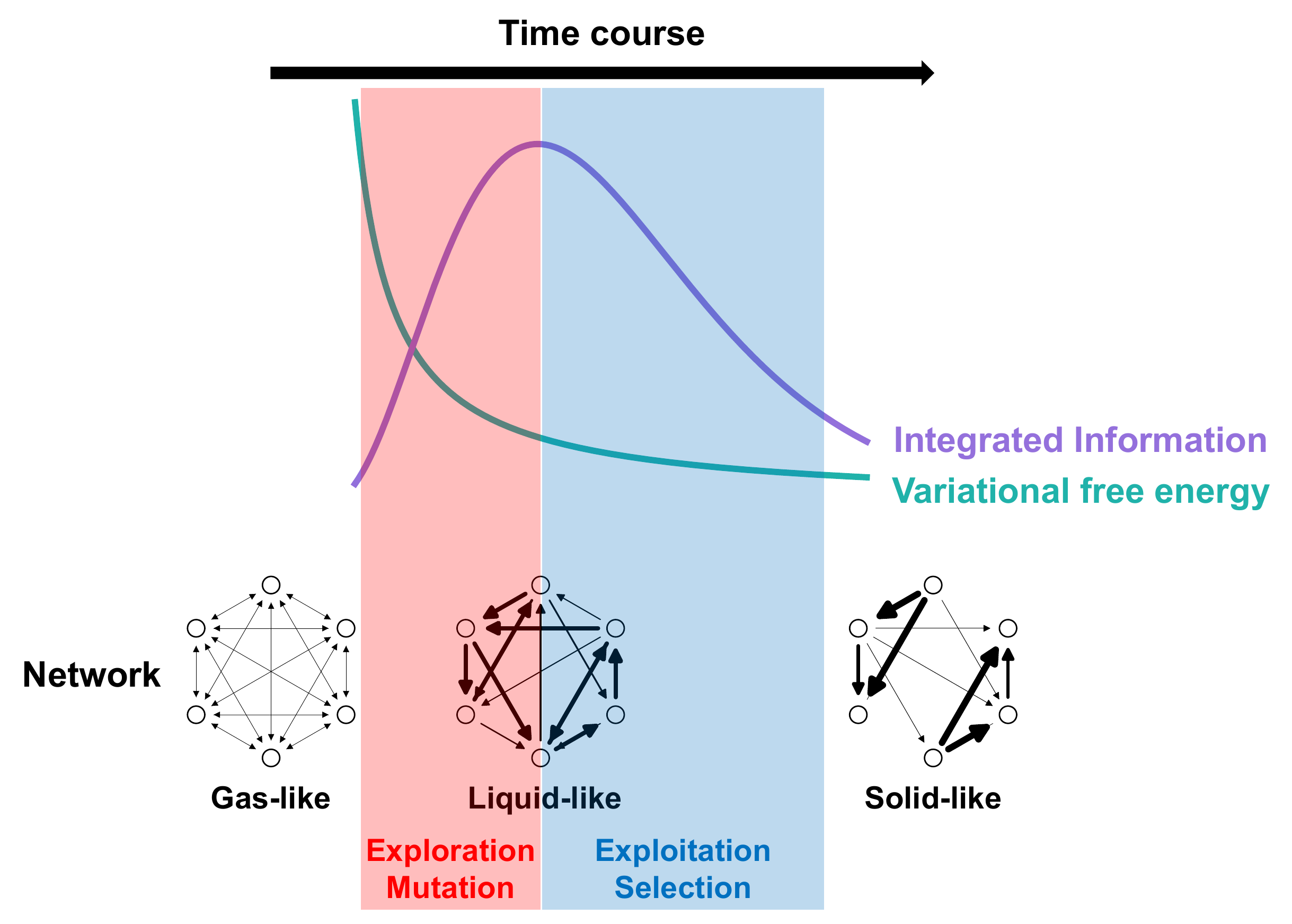}
  \captionsetup{
  font={small,stretch=1.0},
  labelfont=bf,
  justification=raggedright,
  singlelinecheck=false,
  skip=4pt
  }
  \caption{\textbf{Proposed framework.} Schematic of the proposed behavior of integrated information during self-organization under the FEP.
  This schematic is intended as a heuristic and speculative interpretation, not as direct evidence.
  As VFE decreases over time, integrated information is hypothesized to follow a hill-shaped trajectory, reaching a maximum at intermediate stages characterized by complex recurrent connectivity.
  This trajectory can be interpreted as a transition from less structured to more constrained network organization, paralleling exploration versus exploitation and mutation versus selection in Darwinian dynamics.
  The previously reported decrease in surprise with increasing $\Phi$ over evolutionary timescales by Lundbak Olesen et al. \cite{lundbakolesen2023phi} may reflect the ascending phase of this trajectory.
  }
  \label{fig:fig6}
\end{figure}

\section{Discussion}

\subsection{Summary of main findings}
We investigated how integrated information ($\Phi_R^{\text{mc}}$ and coreness) behaves when cultured cortical networks perform perceptual inference formalized under variational Bayes, or the
free-energy principle (FEP). 
Consistent with prior work \cite{isomura2015cultured, isomura2020reverseengineering, isomura2023experimental}, repeated presentation of observations generated by hidden sources elicited robust source selectivity, demonstrating the emergence of perceptual inference in \textsl{in vitro} neuronal networks (Fig.\ref{fig:fig2}).
Session-wise analyses showed that variational free energy (VFE) decreased, while Bayesian surprise (complexity) and accuracy increased, consistent with self-organization under the FEP
(Fig.\ref{fig3:B}--\ref{fig3:D}). 

At the network level, $\Phi_R^{\text{mc}}$ and main-complex size followed a hill-shaped trajectory across sessions and were positively correlated (Fig.\ref{fig4:B}, \ref{fig4:D}--\ref{fig4:F}).
Qualitatively similar $\Phi_R^{\text{mc}}$ trajectories were observed across individual temporal scales (Fig.\ref{fig4:C}).
Informational cores concentrated diverse neuronal activity: mean coreness positively correlated with the session-wise interquartile range (IQR) of evoked responses, and the mean IQR of electrodes inside the main complex consistently exceeded that of electrodes outside (Fig.\ref{fig4:G}--\ref{fig4:I}).

Across experiments, $\Phi_R^{\text{mc}}$ correlated strongly and positively with Bayesian surprise and showed more modest, heterogeneous positive correlations with accuracy and VFE (Fig.\ref{fig5:A}--\ref{fig5:F}).
The exploratory regression analyses further suggest that Bayesian surprise is a more consistent statistical predictor of $\Phi_R^{\text{mc}}$ than accuracy, including after session progression is entered as a covariate (Supplementary Figs.~S1--S3 and Supplementary Table~S1).
Moreover, the spatial allocation of belief updating was associated with the geography of informational cores: coreness contrasts mirrored source-specific Bayesian surprise contrasts (Fig.\ref{fig5:G}, \ref{fig5:H}).
Taken together, these findings suggest that integrated information covaries with belief updating during FEP-consistent learning, rather than directly reflecting model performance.
In this way, our results provide empirical alignment between IIT and the FEP by linking an IIT-inspired measure of intrinsic integrated information with the adaptive dynamics of inference.

\subsection{Integrated information and Bayesian surprise}
A central observation is the robust positive association between $\Phi_R^{\text{mc}}$ and Bayesian surprise across sessions and experiments (meta-analytic Spearman's $\rho=0.886$; Fig.\ref{fig5:C}, \ref{fig5:D}).
Bayesian surprise quantifies the divergence between the prior $P(s)$ and the variational posterior $Q(s) \approx P(s \mid o)$, i.e., the degree of belief updating elicited by new evidence.
When belief change is small---because the current generative model already explains inputs with high likelihood---processing can rely on pre-existing, localized, relatively reflexive circuits with lower irreducibility.
By contrast, when belief change is large, the model must be reconstructed, potentially yielding distributed and synergistic activity patterns that span subnetworks and increase integrated information.
This mechanistic picture is compatible with the session-wise increase in response diversity (IQR) alongside Bayesian surprise (mean $\rho = 0.777$; Fig.\ref{fig3:F}, \ref{fig3:G}), the positive association between main-complex size and $\Phi_R^{\text{mc}}$ (mean $\rho = 0.286$; Fig.\ref{fig4:E}, \ref{fig4:F}), and the consistently higher IQR inside than outside the main complex (Fig.\ref{fig4:G}--\ref{fig4:I}).
Thus, diverse and widely coupled dynamics may accompany belief revision and covary with $\Phi$.
Functionally, because sustaining large $\Phi$ may entail spatial and metabolic costs, it may emerge most clearly when these costs are offset by informative inputs, as reflected in high Bayesian surprise.

Response diversity, quantified as the session-wise IQR of evoked responses, tracked Bayesian surprise across experiments (mean $\rho = 0.777$; Fig.\ref{fig3:F}, \ref{fig3:G}).
This result is compatible with operation near criticality \cite{beggs2003neuronal, pasquale2008self, shew2011information, yada2017development, ikeda2023noise}, a regime in which neural systems have been proposed to maximize dynamic range and stimulus--response mutual information, $I(S{;}R)$, while exhibiting rich long-range correlations.
When $Q(s)$ approximates $P(s \mid o)$ and is averaged over observations, Bayesian surprise relates to the mutual information between observations and hidden states, $I(o{;}s)$.
Because observations $o$ correspond to stimuli and beliefs about hidden states $s$ are encoded in neural responses, increases in $I(S{;} R)$ near criticality could enhance both $I(o{;}s)$ and Bayesian surprise.
Given theoretical and empirical predictions that integrated information $\Phi$ peaks near criticality \cite{aguilera2019integrated, kim2019criticality, popiel2020emergence, mediano2022integrated}, together with our findings of positive IQR--coreness covariation and consistently higher IQR inside than outside the main complex (Fig.\ref{fig4:G}--\ref{fig4:I}), the observed positive correlation between $\Phi$ and Bayesian surprise may share a basis in response diversity and coordinated differentiation.
We did not directly test criticality, however, and therefore treat this interpretation as a hypothesis for future work.

Importantly, Bayesian surprise accords with the intrinsicality emphasized by IIT.
Integrated information structure is fundamentally intrinsic---as in dreaming---but can be modulated by external stimuli.
Bayesian surprise is defined solely in terms of internal elements, the prior $P(s)$ and the variational posterior $Q(s)$, yet depends implicitly on external observations $o$ through $Q(s) \approx P(s \mid o)$.
This perspective aligns with IIT’s claim that integrated information reflects meaningful intrinsic cause--effect power, rather than the extrinsic Shannon-style messages or codes \cite{zaeemzadeh2024shannon}. 

Under the IIT assumption that $\Phi$ underlies consciousness, its coupling with Bayesian surprise may offer a speculative way to think about experiential phenomena in which incoming evidence drives substantial model updating.
For example, in motor adaptation, such as learning to play an instrument, early practice may involve vivid and effortful sensations because internal models are still being reorganized, whereas fluent performance after learning may require less belief updating.
A similar logic may apply to perceptual adaptation, as in glare adjustment or Troxler fading, where phenomenology changes as predictions become more stable.
The same perspective may also help interpret why much spontaneous neural activity is not consciously experienced: such fluctuations may elicit little belief updating and may fail to recruit large integrated structures.
These examples are speculative: our in vitro preparation does not assess subjective experience or consciousness directly.
Rather, the present data suggest that, if integrated information is interpreted in the IIT sense, its relationship to Bayesian surprise may be relevant to future frameworks linking belief updating and phenomenology.

\subsection{Integrated information and accuracy}
Overall, accuracy showed a significant but heterogeneous positive relationship with $\Phi_R^{\text{mc}}$ (mean $\rho = 0.378$; Fig.\ref{fig5:E}, \ref{fig5:F}).
This suggests that greater integration can accompany better inference performance, yet high accuracy is not strictly contingent on large $\Phi_R^{\text{mc}}$: 7/27 experiments exhibited negative correlations.
These results are consistent with the view that rich $\Phi$--structures may confer functional advantages \cite{albantakis2014evolution, grasso2021causal}, while also aligning with IIT's prediction that functionally equivalent systems can differ in their integrated causal structure \cite{oizumi2014from, albantakis2023integrated}.
This dovetails with our observation that $\Phi$ is more consistently associated with belief updating (complexity) than with performance per se.
Three analogies illustrate this dissociation: Bayesian surprise vs. accuracy, model parameter count vs. performance, and intrinsic integrated information vs. extrinsic functionality.
In each case, the former can contribute to the latter, but it is not strictly required.

\subsection{Integrated information and variational free energy}
Empirically, the $\Phi_R^{\text{mc}}$--VFE relationship was positive overall but modest and highly heterogeneous across experiments (mean $\rho = 0.360$; Fig.\ref{fig5:A}, \ref{fig5:B}).
To reconcile this heterogeneous positive association with reports that $\Phi$ increases as surprise falls over longer (evolutionary) timescales \cite{lundbakolesen2023phi} and with theoretical accounts suggesting that minimizing VFE may entail maximizing $\Phi$ \cite{friston2020sentience, safron2020an}, we propose a hill-shaped trajectory: as VFE decreases, $\Phi$ initially rises, peaks, and then declines (conceptual diagram Fig.\ref{fig:fig6}).
This scheme accords with the observed hill-shaped transitions of $\Phi_R^{\text{mc}}$ and main-complex size (Fig.\ref{fig4:B}--\ref{fig4:D}).
Under such a trajectory, $\Phi$--VFE correlations can be positive or negative, depending on whether the system resides on the ascending or descending slope.
This framework accommodates the variability observed across experiments while situating the negative relations reported in theory \cite{friston2020sentience, safron2020an} and in evolutionary simulations \cite{lundbakolesen2023phi} within the ascending phase, without contradicting the modest positive average observed here.

At a high VFE (a maladapted regime), the entropy of observations tends to be large---under ergodicity, the long-term average of VFE serves as an upper bound on observation entropy \cite{friston2013life}---so behavior becomes weakly structured and elements act almost independently.
Integrated information is presumably low owing to the absence of the cause--effect power emphasized by IIT---conceptually, a high-entropy ``gas-like'' network. 
At a very low VFE (an idealized limit), the agent's generative model would predict perfectly and processing would become reflexive and feedforward, with minimal belief updating.
Integrated information should again be low, both because of the spatial and metabolic cost of maintaining it, the absence of recurrence, reduced susceptibility and stereotyped responses---conceptually, a low-entropy ``solid-like'' network.
Between these extremes, the model is competent yet uncertainty remains.
Multiple competing hypotheses must be coordinated and revised by ongoing input, fostering large recurrent cause--effect structures and high $\Phi$---conceptually, a medium-entropy ``liquid-like'' network.

Functionally, this hill-shaped trajectory can be interpreted as a progression from an exploration-like (mutation-like) phase to an exploitation-like (selection-like) phase.
Early in training, high-$\Phi$ states may coincide with information harvesting---high Bayesian surprise, related to the mutual information $I(o{;}s)$---so substantial resources are invested to construct large integrated cores and explore models capable of explaining the inputs with sufficient likelihood.
Later, as the model compresses and stabilizes, exploitation may dominate: $\Phi$ subsides while VFE continues to decline.
A similar interpretation applies to mutation--selection metaphors in neural Darwinism-like dynamics \cite{edelman1993neural, kilgard2012harnessing, takahashi2013response}: early training expands the responsive area and diversifies neural responses (presumably higher $\Phi$), whereas later training contracts the area and stereotypes responses (lower $\Phi$) even as performance improves \cite{takahashi2013response}.
Together, these analyses suggest that $\Phi$ is not a direct proxy for model efficiency.
Instead, it may peak during phases of belief revision embedded within longer-term free-energy descent.

\subsection{Limitations of the present study}
First, the proposed hill-shaped trajectory of $\Phi$ is an idealized principle whose full expression is constrained in practice.
Embodiment, bodily degrees of freedom, and environmental complexity often prolong development, such that a post-developmental state with diminished $\Phi$ may rarely be reached outside of 
simple tasks.
Our \textsl{in vitro}, low-difficulty task with two binary hidden states likely enabled some cultures to reach this exploitative regime. 
Because the preparation was disembodied and passively stimulated, the exploratory stage was probably shorter than would occur in an embodied setting.
In active inference, agents minimize expected free energy, which includes the epistemic-value term (expected Bayesian surprise) with a negative sign \cite{friston2015active, parr2022active}, thereby promoting exploration, sustaining higher Bayesian surprise, and maintaining larger $\Phi$ during active sensing, as in daily active vision \cite{parr2017active}.
Second, $\Phi$ was approximated using $\Phi_R$ \cite{mediano2019beyond} and coreness with a minimum-cut-based method \cite{kitazono2020efficient, kitazono2023bidirectionally}. 
These are IIT-inspired proxies rather than full IIT 3.0/4.0
quantifications.
Our approaches emphasize synergistic coupling but do not exhaustively assess state-dependent cause--effect structures across spatiotemporal scales\cite{albantakis2023integrated}. 
Third, although we conducted the analysis across multiple temporal scales ($\tau=1$--64 ms) and used the $\tau$ that maximized $\Phi_R^{\text{mc}}$ as the primary session-wise metric, the analysis remains scale-dependent.
The explored set of temporal scales was limited, and maximizing across $\tau$ may preferentially select high estimates; accordingly, the fixed-$\tau$ trajectories were treated as a sensitivity check.
We also treated each electrode as a unit and estimated transitions primarily from stimulation (perturbational) trials.
While these approximations are likely reasonable---given the emergence of integrated information at the macro timescale in actual neural recordings \cite{leung2022emergence} and the characteristic timescales of cultured neurons \cite{madhavan2007plasticity, yada2016state}---they warrant cautious interpretation.
Fourth, although our theoretical account invokes positive near-critical modes, we did not directly estimate criticality, avalanche statistics, or the spectrum of the effective interaction matrix from the empirical data.
Finally, substantial between-experiment heterogeneity
(high $Q$, high $I^2$) in several meta-analyses cautions that culture-specific factors (e.g., maturation, connectivity, excitability) may modulate the coupling between $\Phi$, Bayesian surprise, and performance.

\section{Conclusion}
Our results suggest that, in living neuronal networks performing perceptual inference, integrated information is most consistently associated with belief updating---indexed by Bayesian surprise---while showing weaker and more heterogeneous associations with variational free energy and accuracy.
Informational cores expand and concentrate diverse activity when belief revision is stronger, and a $\Phi$-proxy follows a hill-shaped trajectory across learning sessions, peaking within long-term free-energy descent.
These dynamics are compatible with the possibility that response diversity, belief updating, and integrated information are jointly expressed in regimes related to critical dynamics, although criticality was not directly tested here.
Conceptually, $\Phi$ may reflect the intrinsic integrated structure accompanying system reorganization required to incorporate informative evidence; once the generative model becomes sufficiently complete, $\Phi$ may decline.
Functionally, these results are more consistent with $\Phi$ accompanying model updating than serving as a direct proxy for inference performance.
By situating integrated information within belief updating, our findings provide an empirical link between IIT's mechanistic account and the FEP's functional perspective, potentially informing future frameworks that bridge the proximate ``how'' and the ultimate ``why'' of consciousness. 

\section{Methods}

\subsection{Dissociated neuronal cultures}\label{method:culture}
All procedures complied with the “Guiding Principles for the Care and Use of Animals in the Field of Physiological Science” published by the Japanese Physiological Society. 
The Committee on the Ethics of Animal Experiments at the Graduate School of Information Science and Technology, the University of Tokyo, approved the experimental protocol (A2024IST003).

High-density microelectrode arrays (HD-MEAs, MaxOne, MaxWell Biosystems) were covered with 1 mL of 1\% Tergazyme (Sigma-Aldrich) and left at room temperature for 2 h.
The detergent was removed with an aspirator, and the chips were rinsed three times with sterilized water.
Each chip was subsequently soaked in ethanol for 30 min, rinsed three additional times, overlaid with 1 mL of pre-warmed plating medium (Neurobasal Plus, Thermo Fisher Scientific), covered to prevent drying, and maintained in an incubator for at least 2 days.

After this pretreatment, the chips were rinsed three times with sterile water.
Polyethylenimine (Supelco) was diluted to 0.07 \% in borate buffer (Thermo Fisher Scientific), and 50 µL was applied to each electrode surface.
The chips were incubated overnight, washed three times and then coated with 50 µL of laminin (20 µg/mL; Sigma-Aldrich).
After replacing the lids, the chips were incubated for 1 h.

Pregnant Wistar rats were anesthetized with inhaled isoflurane (Viatris) and euthanized by guillotine decapitation.
Following abdominal disinfection with ethanol, the uterus was removed and placed in Hanks’ Balanced Salt Solution (Life Technologies).
Three E18 fetuses were harvested, their brains were removed, and pieces of cerebral cortex were excised for cell seeding.
The sex of the fetuses was not determined.

The cortical tissue was transferred to 2 mL of 0.25 \% Trypsin-EDTA (Thermo Fisher Scientific) and incubated for 20 min, with the tube shaken every 5 min.
The tissue was then transferred to plating medium to stop the enzymatic reaction, gently shaken, and placed in fresh medium. 
Cells were dissociated with trituration by pipetting. 
One milliliter of the suspension was passed through a 40 µm cell strainer (Falcon). Plating medium was added to adjust the density to 38,000 cells per 5 µL.

The laminin solution was removed from the chip surface, and 50 µL of the cell suspension was applied onto the electrodes.
The chip was incubated for 120 min to allow cell attachment, after which 0.6 mL of plating medium was added.
The chip was then maintained in the incubator.
To prevent evaporation, the chip was covered with its lid, placed with a 35 mm dish of sterilized water inside a 90 mm dish, and kept in an incubator at 36.5 °C in 5 \% CO2.

In this study, 12 independent cell cultures were used to conduct 27 experiments.
The average days in vitro (DIV) was $18.4 \pm 6.96$. 

\subsection{Electrophysiological experiments}\label{method:electrophysiology}
HD-MEAs were used both to record the activity of cultured neuronal networks and to deliver electrical stimulation.
The HD-MEA employed in this study contained 26,400 electrodes arranged within an area of 3.85 mm × 2.10 mm, with 17.5-µm spacing between electrodes, of which up to 1,024 could be recorded simultaneously at a sampling rate of 20 kHz \cite{ballini20141024, muller2015high}. 
Prior to experiments, spontaneous activity was recorded from all electrodes for 50 s.
Based on the average spike amplitude during this period, up to 1,024 electrodes with the highest amplitudes were selected for subsequent recordings.
From this set, the 32 electrodes with the highest average spike amplitudes were designated as stimulation electrodes.
Among them, the 16 electrodes with odd-numbered amplitude ranks delivered stimulation corresponding to observations $o^{(1)}-o^{(16)}$, while the 16 electrodes with even-numbered ranks delivered stimulation corresponding to observations $o^{(17)}-o^{(32)}$.
Because recordings from the stimulation electrodes were prone to noise interference, subsequent recordings were obtained from up to 992 electrodes, excluding these 32 stimulation electrodes.
Electrical stimulation consisted of biphasic pulses with a positive-first phase, an amplitude of 350 mV, and a pulse width of 200 $\mu$s. 

\subsection{Data processing}\label{method:data_processing}

For spike detection, the recorded potentials were band-pass filtered (300--3000 Hz, Butterworth filter).
A spike was detected when the measured potential at an electrode fell below a threshold set at five times the standard deviation of the potential for that electrode.

In our samples, spike counts peaked within 100--200 ms after electrical stimulation (Fig.\ref{fig2:B}).
Accordingly, the evoked response strength  $r_{ti}$ at an electrode $i$ in a trial $t$ was defined as the number of spikes occurring within a 10--300 ms window post-stimulation.

This treatment of evoked responses closely followed that of previous studies \cite{isomura2015cultured, isomura2023experimental}; readers are referred to those works for further details.
For trials in which the source state was $(s^{(1)}, s^{(2)}) = (1, 0)$ (approximately 6,400 trials ($=100\text{ sessions}\times256\text{ trials}/\text{session}/4\text{ states}$)), the mean $r_{ti}$ was computed, as well as for trials in which $(s^{(1)}, s^{(2)}) = (0, 1)$ ($\sim$6,400 trials).
The difference between these two means was then calculated for each electrode.
Electrodes with differences $>0$ were classified as $s^{(1)}$-preferring, those with differences $<0$ as $s^{(2)}$-preferring, and those with differences $=0$ as non-preferring/inactive.
The numbers of $s^{(1)}$-preferring, $s^{(2)}$-preferring, and non-preferring/inactive electrodes were $352.0 \pm 359.3$, $353.1 \pm 347.5$, and $287.7 \pm 311.0$, respectively ($n = 27$).

For each trial, the mean evoked response over the $s^{(1)}$-preferring electrodes was computed as $x_{t1}$, and the mean over the $s^{(2)}$-preferring as $x_{t2}$. 
Both $x_{t1}$ and $x_{t2}$ were then mean-subtracted, detrended, and normalized to the range $[0, 1]$.

\subsection{KLD of neuronal response}\label{method:response_KLD}

To evaluate the source selectivity of neuronal responses at each electrode, we used the Kullback-Leibler divergence (KLD) method introduced in a previous study \cite{isomura2015cultured}.
For electrode $i$, the distributions of evoked spike counts in $(s^{(1)},s^{(2)})=(1,0)$ and $(0,1)$ trials were each fitted with a Poisson distribution. 
The empirical parameters $\lambda_{1,0}$ and $\lambda_{0,1}$ were estimated, and the KLD was computed according to the following equation:

$$
D_{\text{KL}}(P(r_{i}\mid(1,0)) \parallel P(r_{i}\mid(0,1))) = \lambda_{1,0}\,\ln \frac{\lambda_{1,0}}{\lambda_{0,1}} + \lambda_{0,1} - \lambda_{1,0}.
$$

In the Results, we report analyses restricted to the 7,613 electrodes for which the computed KLD values converged (i.e., did not diverge).

\subsection{FEP-based analysis}\label{method:FEP}
For FEP-based analysis, we closely followed the methods described in previous studies \cite{isomura2020reverseengineering, isomura2022canonical, isomura2023experimental}, including the generative process, variational Bayesian inference, the canonical neural network, and the reverse-engineering framework.
For mathematical details, readers are referred to those prior studies.

\subsubsection{Generative process of observations}

We assumed a partially observable Markov decision process (POMDP) in which two independent binary hidden sources $s_t = (s^{(1)}_t, s^{(2)}_t) \in \{0,1\}^2$, generated 32 binary sensory observations, $o_t \in \{0,1\}^{32}$, via a stochastic mixing matrix $A$.
In the actual experiment, the state of each hidden source was drawn independently from a Bernoulli distribution with probability 0.5.
For each observation channel, the observation was generated from the hidden sources with specific conditional probabilities.
In particular,  $o^{(1)}-o^{(16)}$ conveyed the value of $s^{(1)}$ with probability 0.75 or that of $s^{(2)}$ with probability 0.25; conversely, $o^{(17)}-o^{(32)}$ conveyed the value of $s^{(2)}$ with probability 0.75 or that of $s^{(1)}$ with probability 0.25.
This defined the categorical likelihood $P(o_t^{(i)} \mid s_t, A)$ for each electrode, with $P(A^{(i)})$ assigned a Dirichlet prior.

\subsubsection{Variational free energy}

Under a mean-field approximation $Q(s_{1:t}, A) = Q(A) \prod_{\tau=1}^t Q(s_\tau)$, the variational free energy (i.e., the negative evidence lower bound) is given by

$$
F = \sum_{\tau=1}^t s_\tau \cdot \big( \ln s_\tau - \ln A \cdot o_\tau - \ln D \big) + O(\ln t),
$$
where $D$ is the prior over hidden states.
Minimizing $F$ with respect to $s_\tau$ and the Dirichlet parameters $a$ yields

$$
s_\tau = \sigma \!\big( \ln A \cdot o_\tau + \ln D \big), \quad a \leftarrow a + \sum_{\tau=1}^t o_\tau \otimes s_\tau,
$$
where $\sigma(\cdot)$ is the softmax function and $\otimes$ denotes the outer product.

\subsubsection{Canonical neural network formulation}

Neuronal responses $x_t \in (0,1)^2$ to sensory inputs $o_t$ were modeled as a canonical neural network with the following dynamics:

$$
\dot{x}_t \propto -\mathrm{sig}^{-1}(x_t) + W o_t + h,
$$
where $\mathrm{sig}^{-1}(\cdot)$ is the elementwise logit function, $W$ is a $2 \times 32$ synaptic strength matrix, and $h$ is the adaptive firing threshold vector.
The matrix $W = W_1 - W_0$ is composed of excitatory ($W_1$) and inhibitory ($W_0$) components.

\subsubsection{Neural network cost function $L$}

Integrating the network dynamics with respect to $x_t$ yields a cost function

$$
L = \sum_{\tau=1}^t 
\begin{pmatrix} x_\tau \\ \bar{x}_\tau \end{pmatrix}^\top 
\Bigg[
\ln \begin{pmatrix} x_\tau \\ \bar{x}_\tau \end{pmatrix}
- 
\ln \begin{pmatrix} \hat{W_1} & \bar{\hat{W_1}} \\ \hat{W_0} & \bar{\hat{W_0}} \end{pmatrix}
\begin{pmatrix} o_\tau \\ \bar{o}_\tau \end{pmatrix}
-
\begin{pmatrix} \phi_1 \\ \phi_0 \end{pmatrix}
\Bigg]
+ C,
$$
where $\bar{x} = 1 - x$, $\bar{o} = 1 - o$, $\hat{W}_\ell = \mathrm{sig}(W_\ell)$, and $\bar{\hat{W_\ell}} = 1 - \mathrm{sig}(W_\ell)$. 
The threshold factors $\phi = (\phi_1, \phi_0)^\top$ correspond to $\ln{D}$.
This $L$ is asymptotically equivalent to $F$, with $x \leftrightarrow s$, $W \leftrightarrow A$, and $\phi \leftrightarrow \ln D$.

\subsubsection{Reverse engineering from empirical neural activity}

From experimental data, neuronal responses $x_t$ were calculated for each trial.
Given these responses, the threshold factor $\phi$ was then estimated as:

$$
\phi = \begin{pmatrix} \phi_1 \\ \phi_0 \end{pmatrix} = \ln \begin{pmatrix} \langle x \rangle \\ 1 - \langle x \rangle \end{pmatrix},
$$
where $\langle \cdot \rangle$ indicates the average over time.
The threshold factor $\phi$ was held constant within each session.
Following previous studies, $\phi$ for the first 10 sessions was computed as the average of the neuronal responses during those sessions.
For subsequent sessions, $\phi$ was computed as the average of the neuronal responses in the immediately preceding session.

The effective synaptic connectivity $W$ was estimated from the outer products of $x_t$ and $o_t$ according to the fixed-point equations

$$
W_1 = \mathrm{logit} \!\left( \frac{\langle x o^\top \rangle}{\langle x \mathbf{1}^\top \rangle} \right), \quad
W_0 = \mathrm{logit} \!\left( \frac{\langle (1-x) o^\top \rangle}{\langle (1-x) \mathbf{1}^\top \rangle} \right), \quad W = W_1 - W_0.
$$

Substituting $x$, $W$, and $\phi$ into $L$ yielded the empirical variational free energy for each session.
At the same time, we computed empirical Bayesian surprise
$$
\sum_{\tau=1}^t 
\begin{pmatrix} x_\tau \\ \bar{x}_\tau \end{pmatrix}^\top 
\Bigg[
\ln \begin{pmatrix} x_\tau \\ \bar{x}_\tau \end{pmatrix}
-
\begin{pmatrix} \phi_1 \\ \phi_0 \end{pmatrix}
\Bigg]
$$
and empirical accuracy
$$
\sum_{\tau=1}^t 
\begin{pmatrix} x_\tau \\ \bar{x}_\tau \end{pmatrix}^\top 
\ln \begin{pmatrix} \hat{W_1} & \bar{\hat{W_1}} \\ \hat{W_0} & \bar{\hat{W_0}} \end{pmatrix}
\begin{pmatrix} o_\tau \\ \bar{o}_\tau \end{pmatrix}.
$$

\subsection{Neuronal response IQR}\label{response_IQR}

To evaluate the variability of neuronal responses, we used the interquartile range (IQR). 
For each session, the mean evoked response $r^{(1)}$ of $s^{(1)}$-preferring electrodes was grouped by hidden source state, and the IQR was calculated within each group. 
These IQR values were then averaged. 
The same procedure was applied to $r^{(2)}$ of $s^{(2)}$-preferring electrodes. 
The two resulting IQRs were then averaged to yield an overall measure of response diversity for the network.

Similarly, to assess the variability of neuronal responses at a single electrode, trials were grouped by hidden source state, and the IQR was calculated within each group and then averaged.

\subsection{Transition probability}\label{method:transition_probability}

In IIT, the cause--effect power is evaluated from the transition probabilities between system states.
The method used here corresponds to what has previously been referred to as the downsampling method \cite{leung2022emergence}.
Specifically, the time series was coarse-grained into states by segmenting it into windows of width $\tau$, and the empirical distribution of state transitions between adjacent windows was computed.
For a time series of length $T$, there are $T - \tau + 1$ such windows, each represented by the mean value of the observations within that window.
These representative values were binarized using their median as the threshold.

Adjacent pairs of windows yield $T - 2 \tau + 1$ transitions, which were used to compute state transition probabilities.
In each trial, evoked responses during 10--300 ms after stimulation were binned at 1-ms resolution, resulting in a time series of length 290.
For each session, a single state transition probability matrix was computed using all trials in which electrical stimulation was delivered, i.e., those with $(s^{(1)}, s^{(2)}) \neq (0, 0)$, amounting to approximately 256×3/4=192 trials. 
The use of only trials containing stimulation followed the rationale of the perturbational approach.
\subsection{Complex extraction}\label{method:complex_extraction}

For each session, a weighted undirected graph was constructed in which each vertex represented a preferring electrode, and all vertices were fully connected.
The weight of each edge was given by $\Phi_R$ \cite{mediano2019beyond}, computed from the neuronal activity of the corresponding pair.
The number of vertices occasionally reached $\sim$900.
The state transition probabilities were calculated for all electrode pairs across multiple temporal scales ($\tau=1,2,4,8,16,32,$ and $64$ ms), and $\Phi_R$ values were derived from these transition probabilities. 
The $\Phi_R$ between electrodes $X$ and $Y$ was expressed as:
$$
\Phi_R(X,Y) = I(X_{t-1}, Y_{t-1}; X_{t}, Y_{t}) - I(X_{t-1}; X_{t}) - I(Y_{t-1};Y_{t}) + \min_{Z = X, Y, W = X, Y} I(Z_{t-1}; W_{t}), 
$$
where $I$ is Shannon's mutual information, and the fourth term corresponds to the minimum mutual information (MMI) \cite{barrett2015exploration} redundancy function, introduced as a corrective measure to avoid negative values.

The method of complex extraction followed that described in previous research \cite{kitazono2020efficient}, and mathematical details are provided therein.
The graph was recursively partitioned using the minimum cut (mc) until all vertices were isolated.
Given a vertex set, the minimum cut is defined as the bipartition of the set into two non-empty, disjoint subsets that minimizes the sum of the edge weights crossing the partition.
The sum of these edge weights crossing the minimum cut of a vertex set was denoted $\Phi_R^{\text{mc}}$ for that set.
A vertex set was defined as a complex if its $\Phi_R^{\text{mc}}$ was greater than that of any of its supersets, and, among such complexes, was further defined as a main complex if its $\Phi_R^{\text{mc}}$ was not smaller than that of any of its subsets.
For each session and temporal scale $\tau$, the maximum $\Phi_R^{\text{mc}}$ among main complexes --- analogous to the integrated information quantity in IIT 2.0 --- was computed.
Then $\Phi_R^{\text{mc}}$ was normalized by the number of edges in the graph. 
This adjustment was necessary because, for two graphs with comparable average edge weights but different numbers of vertices, the graph with more vertices and edges would naturally yield a larger number of edges crossing a cut, and thus a larger $\Phi_R^{\text{mc}}$. 
Normalization by edge count therefore enabled comparisons across graphs of different sizes.
Although multiple values of $\Phi_R^{\text{mc}}$ corresponding to different $\tau$ values were obtained for each session, we used the maximal $\Phi_R^{\text{mc}}$ value across $\tau$ as the primary session-wise metric, in keeping with the IIT motivation that integrated information should be measured at the spatiotemporal scale where it is maximized.
Fixed-$\tau$ trajectories were retained as sensitivity analyses.
The size of the main complex was computed at the $\tau$ that maximized $\Phi_R^{\text{mc}}$ for that session.

Additionally, we computed the coreness measure \cite{kitazono2023bidirectionally}.
For a given graph, the coreness of a vertex is defined as the maximum $\Phi_R^{\text{mc}}$ among all complexes that include that vertex (noting that the set of all vertices always constitutes at least one complex).
In previous work \cite{kitazono2023bidirectionally}, coreness was computed for the mouse connectome and found to be high in regions such as the cerebral cortex, which are conducive to large integrated information, and low in regions such as the cerebellum, which are less suited for integrated information.
Thus, coreness quantifies the contribution of each vertex to the system’s integrated information.
Coreness was also computed at the temporal scale $\tau$ that maximized $\Phi_R^{\mathrm{mc}}$ for that session.

\subsection{Regression analysis}\label{method:regression}
To examine the dependency structure among Bayesian surprise (BS), accuracy (Acc), and $\Phi_R^{\text{mc}}$, we fitted within-experiment linear regression models across sessions:
$$
\Phi_R^{\text{mc}} \sim BS, \quad \Phi_R^{\text{mc}} \sim Acc, \quad \Phi_R^{\text{mc}} \sim BS + Acc.
$$
To assess whether these associations were attributable to monotonic session progression, the same models were also fitted with session index $t$ as an additional covariate.
Variables were z-scored within each experiment.
Experiment-wise coefficients were summarized using random-effects meta-analysis and are reported as exploratory analyses in the Supplementary Information (Supplementary Figs.~S1--S3 and Supplementary Table~S1).

\subsection{Statistics and reproducibility}\label{method:statistics}

No formal a priori power analysis was performed. 
The final dataset comprised 27 experiments from 12 neuronal cultures. Experiments with severe recording noise that precluded reliable analysis were excluded during quality control before statistical analysis. 
Within the 27 included experiments, all 100 sessions were analyzed, yielding a total of 2,700 sessions; no sessions were excluded.
Blinding was not performed because the analyses required access to the source-state and stimulation labels and were conducted using predefined computational procedures. 
No allocation to experimental groups was involved.

For comparisons between two paired groups, the Wilcoxon signed-rank test was used.
For the meta-analysis of Spearman correlation coefficients $\rho_i$ obtained from each experiment, values were first transformed into the Fisher-$z$ domain: $z_i = \frac{1}{2}\ln\frac{1+\rho_i}{1-\rho_i}$.
Sampling variances were approximated as $\mathrm{Var}(z_i) \approx (1+\rho_i^2/2)/(n-3)$ \cite{bonett2000sample}, where $n$ denotes the number of paired observations (i.e., the number of data points per experiment contributing to the correlation).
Between-experiment heterogeneity was assessed using the $Q$ statistic and the $I^2$ statistic \cite{higgins2002quantifying}.
Given the presence of heterogeneity, we estimated pooled effects using a random-effects model with DerSimonian--Laird estimation \cite{dersimonian1986meta} of the between-experiment variance $\tau^{2}$.
Random-effects weights were defined as $w_i = 1/(\mathrm{Var}(z_i)+\tau^{2})$, and the pooled effect size was computed as $z_{\text{RE}}=\sum_i w_i z_i / \sum_i w_i$.
The corresponding standard error was $\text{SE}_{\text{RE}} = \sqrt{1/\sum_i w_i}$ and p-values were obtained from the two-sided $Z$-test.
Finally, $z_{\text{RE}}$ and the 95\% confidence interval $z_{\text{RE}} \pm 1.96 \, \text{SE}_{\text{RE}}$ were back-transformed to the correlation scale using $\rho=\tanh(z)$.
Random-effects estimates, together with heterogeneity statistics ($Q$, $I^2$, and $\tau^2$), are reported in the Results.
All random-effects meta-analyses included the effect estimates
obtained from the 27 included experiments.

For the meta-analysis of regression coefficients, the coefficient estimate $\hat{\beta}_i$ and its squared standard error $\mathrm{SE}(\hat{\beta}_i)^2$ from each experiment were used as the effect size and sampling variance, respectively.
Coefficients were pooled using the same DerSimonian--Laird random-effects procedure.
Because these regressions were intended to characterize conditional session-wise associations among variables derived from the same recordings, the coefficients were interpreted descriptively rather than causally.

Distributional normality was not assumed. 
Paired comparisons were therefore performed using two-sided Wilcoxon signed-rank tests, and Spearman's rank correlation was used to assess monotonic associations without assuming linearity or normally distributed variables. 
All statistical tests were two-sided, and the significance threshold was
set at $\alpha = 0.05$. 
Exact $P$ values and sample sizes for the principal inferential tests are reported in the Results and figure legends. 
No correction for multiple comparisons was applied. 
For empirical trajectory plots, lines represent the mean across experiments and shaded regions represent the standard error of the mean, unless otherwise stated.

\subsection{Code availability}
All analysis code except for complex extraction is archived at Zenodo \url{https://doi.org/10.5281/zenodo.20625936} \cite{mayama2026code20625936} and is also available at GitHub \url{https://github.com/yunipoke/Bridging_integrated_information_theory_and_the_free_energy_principle_in_living_neuronal_networks}. 
For complex extraction, the original source \url{https://github.com/JunKitazono/BidirectionallyConnectedCores} \cite{kitazono2023bidirectionally} was utilized. 
Code for the variational Bayesian metric was created with significant reference to the original source \url{https://github.com/takuyaisomura/reverse_engineering} \cite{takuyaisomura2023reverseengineering}.

\subsection{Use of AI-assisted tools}
We disclose that AI-assisted tools, including ChatGPT, were used only to help improve the clarity, grammar, and presentation of some text. These tools were not used to generate data, perform analyses, produce figures or multimedia, draw scientific conclusions, or write cited sources. No AI program is listed as an author. All authors have reviewed and take full responsibility for the accuracy, integrity, originality, and appropriate citation of the submitted work.

\backmatter

\bmhead{Acknowledgements}
We are deeply grateful to Drs. Naotsugu Tsuchiya, Masafumi Oizumi, Muneki Ikeda, Francesco Ellia, Matteo Grasso, Shosuke Nishimoto and Takuya Isomura for valuable discussions and insightful comments.

This work was partially supported by JSPS KAKENHI (23H03023, 24K20854, 25H02600, 25K22825, 26H02517), AMED (24wm0625401h0001), the Asahi Glass Foundation, and the Secom Science and Technology Foundation.

\section*{Competing interests}
There are no competing interests to declare.

\section*{Data availability}
Processed spike data and stimulation conditions (2 hidden source states and 32 observations per trial) have been deposited in the DANDI Archive \url{https://doi.org/10.48324/dandi.001611/0.260611.0634} \cite{mayama2026dandi001611}. 
Derivatives (neuronal responses, PSTH, response KLD, preferring electrodes, VFE, Bayesian surprise, accuracy, $\Phi_R$ adjacency matrices, main-complex membership, and coreness) and Source Data are available at Zenodo \url{https://doi.org/10.5281/zenodo.19901678} \cite{mayama2026zenodo19901678} and \url{https://doi.org/10.5281/zenodo.20625764} \cite{mayama2026zenodo20625764}. 

\section*{Author contributions}
Teruki Mayama: Conceptualization (lead); Investigation (lead); Resources (supporting); Software (supporting); Formal analysis (lead); Visualization (lead); Writing -- original draft (lead). Dai Akita: Formal analysis (supporting); Resources (lead); Funding acquisition (supporting); Project administration (supporting); Supervision (supporting); Writing -- review \& editing (supporting). Sota Shimizu: Software (lead); Resources (supporting). Yuki Takano: Software (supporting); Resources (supporting). Hirokazu Takahashi: Funding acquisition (lead); Project administration (lead); Supervision (lead); Writing -- review \& editing (lead).

\end{document}

\begin{center}
{\Large Supporting Information for\par}
\vspace{0.7em}
{\LARGE Bridging integrated information theory and the free-energy principle in living neuronal networks\par}
\vspace{1em}
Teruki Mayama, Dai Akita, Sota Shimizu, Yuki Takano, and Hirokazu Takahashi\par
\vspace{0.5em}
Correspondence to: Hirokazu Takahashi; E-mail: takahashi@i.u-tokyo.ac.jp
\end{center}

\section*{Supplementary regression analyses}

All regression analyses were performed within each experiment using z-scored session-wise variables.
The dependent variable was the session-wise integrated-information proxy $\Phi_R^{\mathrm{mc}}$.
Bayesian surprise (BS), accuracy (Acc), and session index $t$ were used as predictors.
The analyses were intended to characterize conditional associations among variables derived from the same recordings; they should therefore be interpreted descriptively rather than causally.

\begin{table}[!ht]
\centering
\caption{Summary of exploratory regression models.}
\label{tab:S1}
\resizebox{\linewidth}{!}{%
\begin{tabular}{lccccccc}
\toprule
Model & $R^2$ & $\beta_{\mathrm{BS}}$ & $\beta_{\mathrm{BS}}>0$ & $p_{\mathrm{BS}}<0.05$ & $\beta_{\mathrm{Acc}}$ & $\beta_{\mathrm{Acc}}>0$ & $p_{\mathrm{Acc}}<0.05$ \\
\midrule
$\Phi_R^{\mathrm{mc}} \sim \mathrm{BS}$ & 0.735 [0.488, 0.834] & 0.858 [0.695, 0.913] & 27 & 27 & -- & -- & -- \\
$\Phi_R^{\mathrm{mc}} \sim \mathrm{Acc}$ & 0.359 [0.140, 0.627] & -- & -- & -- & 0.473 [$-0.140$, 0.768] & 18 & 22 \\
$\Phi_R^{\mathrm{mc}} \sim \mathrm{BS}+\mathrm{Acc}$ & 0.809 [0.639, 0.900] & 0.829 [0.597, 0.967] & 26 & 26 & $-0.124$ [$-0.242$, 0.272] & 12 & 23 \\
$\Phi_R^{\mathrm{mc}} \sim \mathrm{BS}+t$ & 0.788 [0.711, 0.892] & 0.696 [0.439, 0.956] & 27 & 27 & -- & -- & -- \\
$\Phi_R^{\mathrm{mc}} \sim \mathrm{Acc}+t$ & 0.661 [0.532, 0.859] & -- & -- & -- & 0.135 [$-0.255$, 0.260] & 15 & 19 \\
$\Phi_R^{\mathrm{mc}} \sim \mathrm{BS}+\mathrm{Acc}+t$ & 0.838 [0.742, 0.936] & 0.685 [0.447, 0.857] & 27 & 26 & $-0.106$ [$-0.231$, 0.114] & 10 & 15 \\
\bottomrule
\end{tabular}%
}
\caption*{Values for $R^2$ and regression coefficients are medians[IQR] across 27 experiments. Count columns indicate the number of experiments, out of 27, in which the coefficient was positive or significant at $p<0.05$. ``--'' indicates that the predictor was not included in the model.}
\end{table}

\begin{figure}[!ht]
\centering
\includegraphics[
draft=false,
width=1.0\linewidth,
height=1.0\textheight,
keepaspectratio
]{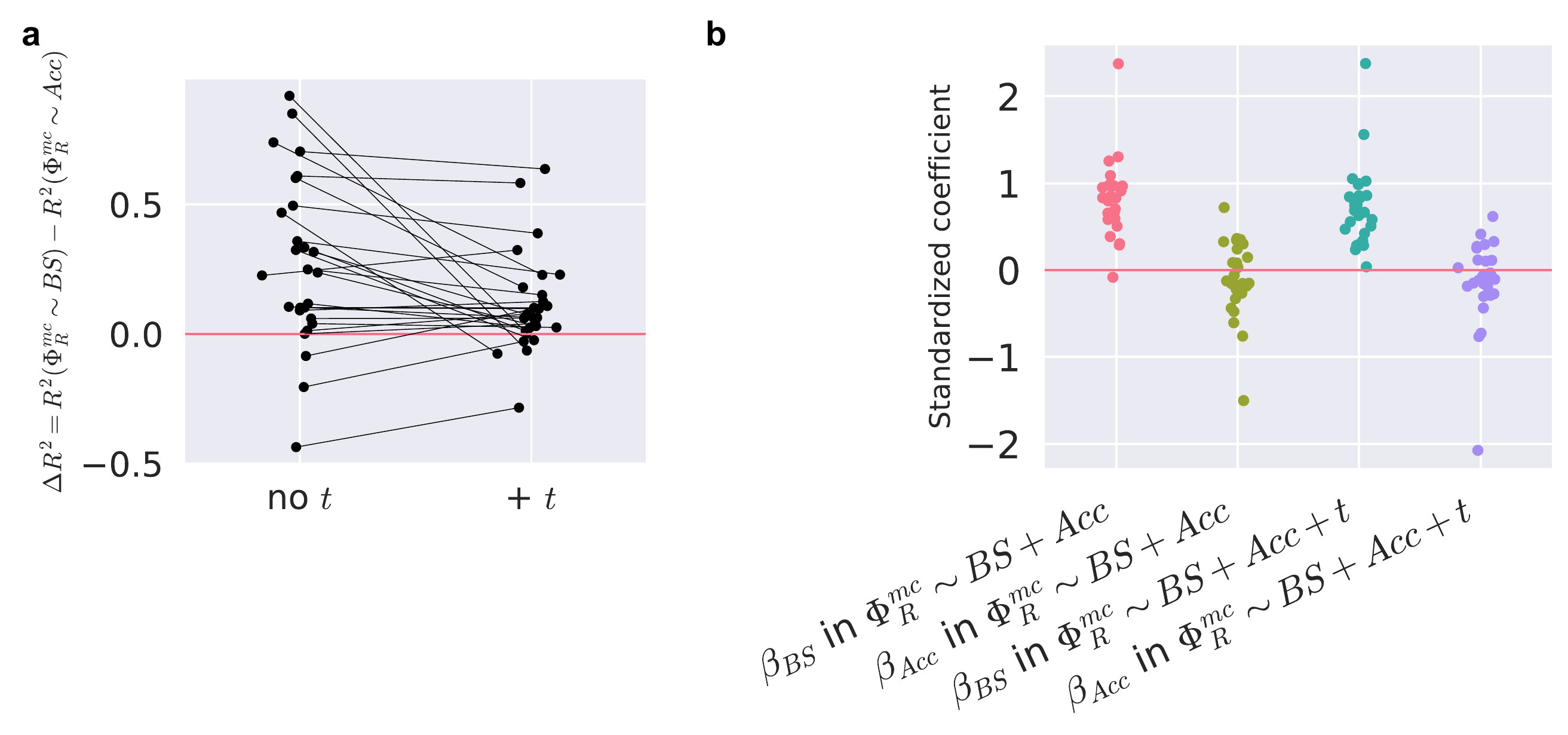}
\captionsetup{
font={small,stretch=1.0},
labelfont=bf,
justification=raggedright,
singlelinecheck=false,
skip=4pt
}
\caption{Comparison of Bayesian-surprise and accuracy regression models.
(a) Difference in coefficient of determination between the Bayesian-surprise model and the accuracy model, computed as $\Delta R^2=R^2(\Phi_R^{\mathrm{mc}}\sim \mathrm{BS})-R^2(\Phi_R^{\mathrm{mc}}\sim \mathrm{Acc})$, without and with session index $t$ as a covariate.
Positive values indicate that Bayesian surprise accounted for more session-wise variation in $\Phi_R^{\mathrm{mc}}$ than accuracy.
(b) Standardized coefficients for Bayesian surprise and accuracy in multivariable models without and with the session-index covariate.
Together with Table~\ref{tab:S1}, these results show that Bayesian surprise was a more consistent statistical predictor of $\Phi_R^{\mathrm{mc}}$ than accuracy.}
\label{fig:S1}
\end{figure}

\begin{figure}[!ht]
\centering
\includegraphics[
draft=false,
width=1.0\linewidth,
height=1.0\textheight,
keepaspectratio
]{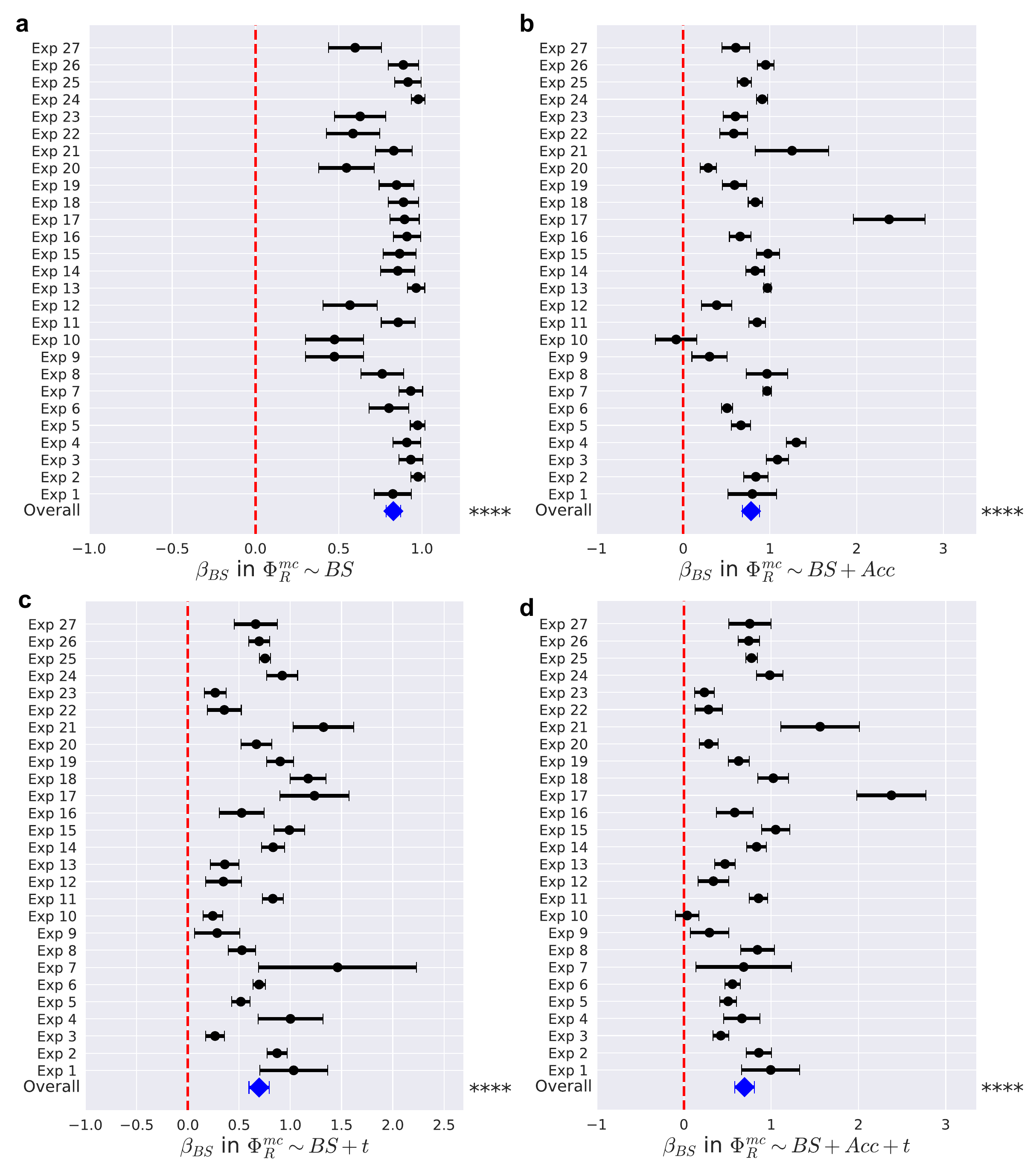}
\captionsetup{
font={small,stretch=1.0},
labelfont=bf,
justification=raggedright,
singlelinecheck=false,
skip=4pt
}
\caption{Random-effects meta-analysis of standardized Bayesian-surprise coefficients.
Panels show $\beta_{\mathrm{BS}}$ for (a) $\Phi_R^{\mathrm{mc}}\sim \mathrm{BS}$, (b) $\Phi_R^{\mathrm{mc}}\sim \mathrm{BS}+\mathrm{Acc}$, (c) $\Phi_R^{\mathrm{mc}}\sim \mathrm{BS}+t$, and (d) $\Phi_R^{\mathrm{mc}}\sim \mathrm{BS}+\mathrm{Acc}+t$.
The pooled Bayesian-surprise coefficient was positive in all models: (a) $\beta=0.829$, 95\% CI [0.785, 0.872], $p=8.4\times10^{-309}$, $Q=177.3$, $\tau^2=0.010$, $I^2=85.3$; (b) $\beta=0.782$, 95\% CI [0.684, 0.880], $p=4.9\times10^{-55}$, $Q=635.9$, $\tau^2=0.061$, $I^2=95.9$; (c) $\beta=0.696$, 95\% CI [0.596, 0.795], $p=6.3\times10^{-43}$, $Q=437.4$, $\tau^2=0.060$, $I^2=94.1$; and (d) $\beta=0.694$, 95\% CI [0.582, 0.806], $p=6.4\times10^{-34}$, $Q=464.2$, $\tau^2=0.078$, $I^2=94.4$.}
\label{fig:S2}
\end{figure}

\begin{figure}[!ht]
\centering
\includegraphics[
draft=false,
width=1.0\linewidth,
height=1.0\textheight,
keepaspectratio
]{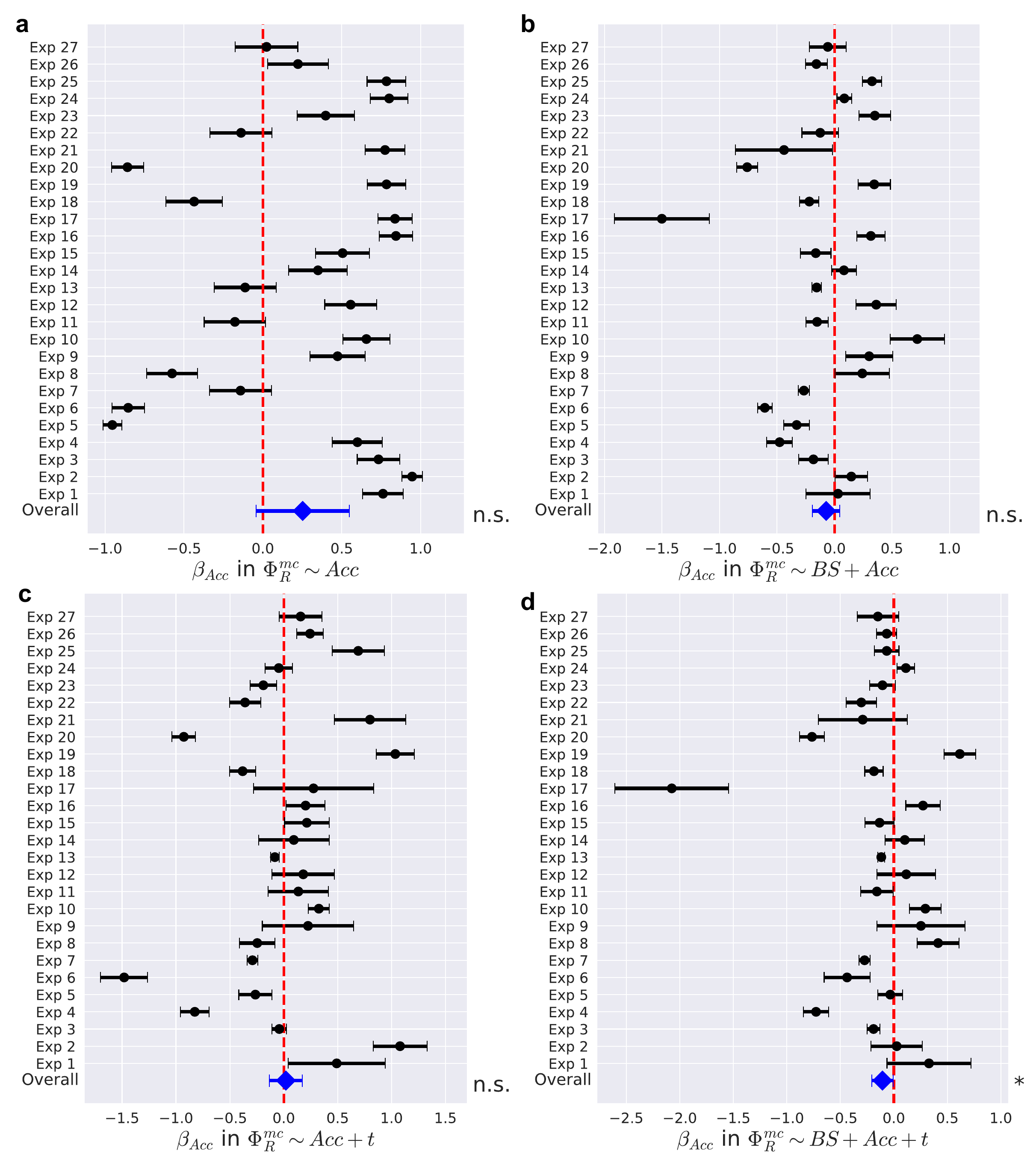}
\captionsetup{
font={small,stretch=1.0},
labelfont=bf,
justification=raggedright,
singlelinecheck=false,
skip=4pt
}
\caption{Random-effects meta-analysis of standardized accuracy coefficients.
Panels show $\beta_{\mathrm{Acc}}$ for (a) $\Phi_R^{\mathrm{mc}}\sim \mathrm{Acc}$, (b) $\Phi_R^{\mathrm{mc}}\sim \mathrm{BS}+\mathrm{Acc}$, (c) $\Phi_R^{\mathrm{mc}}\sim \mathrm{Acc}+t$, and (d) $\Phi_R^{\mathrm{mc}}\sim \mathrm{BS}+\mathrm{Acc}+t$.
The pooled accuracy coefficient was weaker and more heterogeneous than the Bayesian-surprise coefficient: (a) $\beta=0.252$, 95\% CI [$-0.044$, $0.548$], $p=9.5\times10^{-2}$, $Q=328.3$, $\tau^2=0.609$, $I^2=99.3$; (b) $\beta=-0.072$, 95\% CI [$-0.192$, $0.047$], $p=0.235$, $Q=961.2$, $\tau^2=0.093$, $I^2=97.3$; (c) $\beta=0.020$, 95\% CI [$-0.132$, $0.172$], $p=0.799$, $Q=1042.6$, $\tau^2=0.149$, $I^2=97.5$; and (d) $\beta=-0.106$, 95\% CI [$-0.205$, $-0.008$], $p=3.5\times10^{-2}$, $Q=560.8$, $\tau^2=0.059$, $I^2=95.4$.}
\label{fig:S3}
\end{figure}

\begin{figure}[!ht]
\centering
\includegraphics[
draft=false,
width=0.50\linewidth,
height=1.0\textheight,
keepaspectratio
]{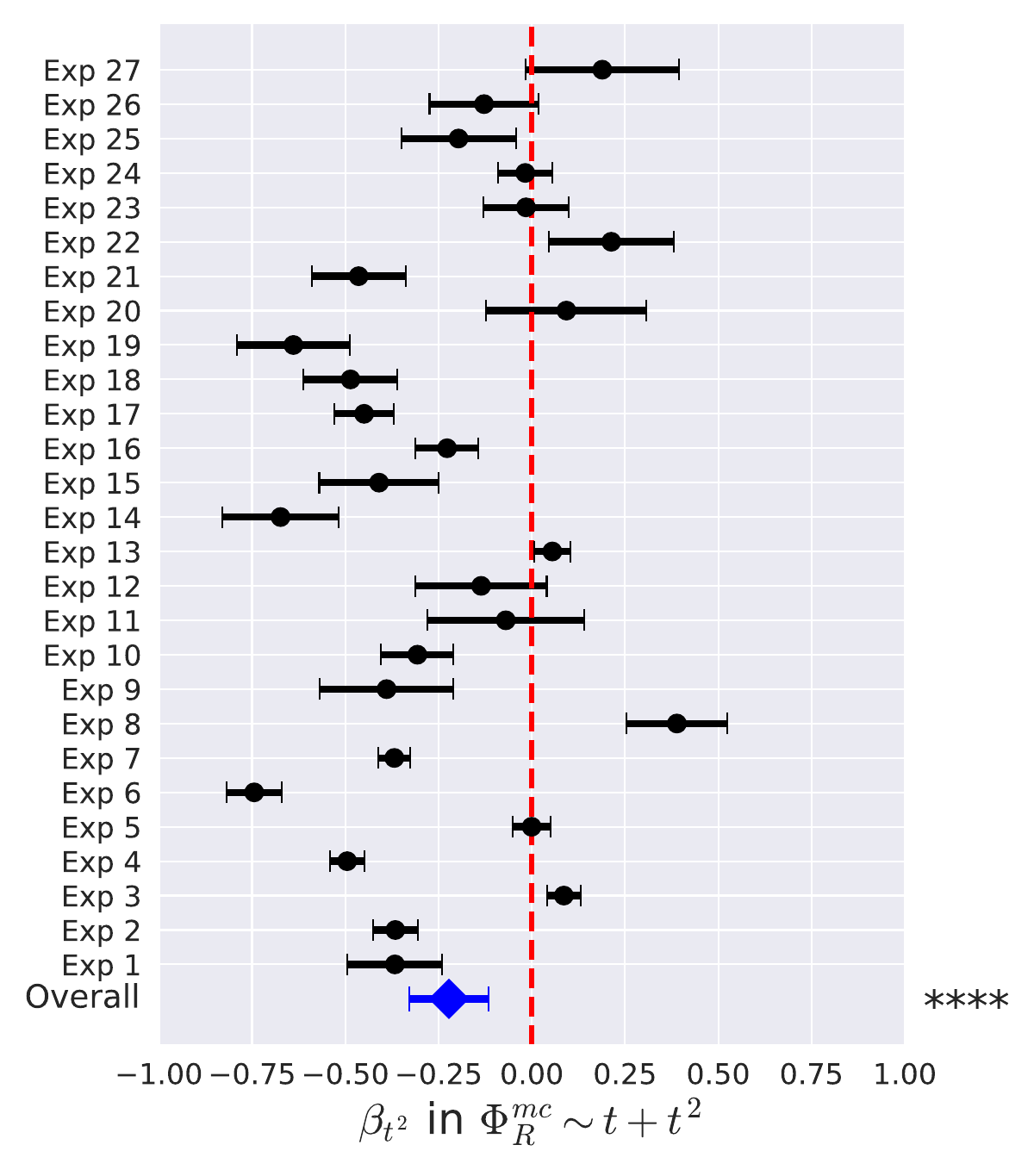}
\captionsetup{
font={small,stretch=1.0},
labelfont=bf,
justification=raggedright,
singlelinecheck=false,
skip=4pt
}
\caption{Random-effects meta-analysis of standardized $t^2$ coefficients.
Random-effects meta-analysis of the squared-session coefficient from within-experiment regressions of z-scored $\Phi_R^\text{mc}$ on standardized session index $t$ and its square $t^2$.
Panel shows $\beta_{t^2}$ for $\Phi_R^{\mathrm{mc}}\sim t+t^2$.
The pooled coefficient is significantly negative: $\beta=-0.222$, 95\% CI [$-0.328$, $-0.117$], $p=3.7\times10^{-5}$, $Q=1084.5$, $\tau^2=0.0739$, $I^2=97.6$. 
The coefficient was negative in $21/27$ experiments and significantly negative in $15/27$ experiments.
The estimated quadratic vertices $-\beta_t/2\beta_{t^2}$ were, after removing the effect of z-scoring of $t$, found to fall between sessions $1$ and $100$ in $17/27$ experiments.
}
\label{fig:S4}
\end{figure}

\clearpage

\section*{Supplementary Note: Theoretical Analysis}
This supplementary note presents a theoretical analysis that provides a possible mechanistic picture of the empirical findings reported in the main text. 
The analysis uses an Ising neural Bayesian model to formalize belief updating and a linear response approach to quantify integrated informational structure.
Its purpose is to show that, under explicit approximations and conditions, Bayesian surprise and a partition-based integrated-information proxy can covary and integrated information can follow a hill-shaped trajectory.

\section*{Ising neural Bayesian model}

We first specify the probabilistic model and the representational assumption connecting neural activity to posterior belief.
The model is based on the Ising model \cite{newell1953theory, brush1967history}, which has been used in statistical mechanics to describe magnetic materials. 
The Ising model has also been applied to neuroscience \cite{schneidman2006weak, roudi2009statistical}. 
Here, we introduce our own Ising neural Bayesian model, in which we equate the activity of a unit with the expected value of the hidden state in the belief distribution.

Let
\[
    S=(S_1,\ldots,S_N), \qquad S_i\in\{-1,+1\},
\]
be a random vector of binary hidden states, and let $s=(s_1,\ldots,s_N)$ denote one realized configuration of $S$.
For a field vector $h\in\mathbb{R}^N$, a symmetric interaction matrix $K\in\mathbb{R}^{N\times N}$, and the state space $\Omega\coloneqq\{-1,1\}^N$, define the partition function
\[
Z(h,K)=\sum_{s\in\Omega}
\exp\left\{h^\top s+\frac{1}{2}s^\top K s\right\}.
\]
The Ising neural Bayesian model assumes that the belief over $S$ is
\[
P_{h,K}(S=s)
=
\frac{1}{Z(h,K)}
\exp\left\{h^\top s+\frac{1}{2}s^\top K s\right\} .
\]
We denote the mean of $S$ under $P_{h,K}$ as $m(h, K)$:
\[
m(h, K) = \mathbb{E}_{P_{h,K}}[S],
\qquad
m_i(h,K)= \mathbb{E}_{P_{h,K}}[S_i] = \sum_{s\in\Omega}s_i P_{h,K}(S=s).
\]
The prior belief and the observation-induced posterior belief are 
\[
    P_p=P_{h_p,K},
    \qquad
    P_o=P_{h(o),K},
\]
where $h_p$ is the prior field and $h(o)$ is the posterior field after observation $o$. 
We assume $K_{ii}=0$. 
The observation is assumed to change only the local field,
\[
    \delta h(o)=h(o)-h_p,
\]
while $K$ is held fixed. 
Thus the observation updates the posterior state by changing local field, not by changing the interaction matrix itself.

The key neural representation assumption is that the mean activity rate of unit $i$ is directly the posterior marginal probability that the corresponding binary hidden state is active:
\[
P_o(S_i=+1) = \frac{m_i(o)+1}{2},
\]
where $m_i(o) = m_i(h(o), K)$.

\section*{Mean-field fixed point and local linearization}

We next approximate the posterior mean and derive the local response around that mean using a mean-field approximation. 
Linearization describes how this fixed point responds to small perturbations.

The exact posterior mean $m(o)=\mathbb E_{P_o}[S]$ is generally difficult to compute for large $N$.
This motivates the use of the well-known mean-field approximation \cite{kappen1998efficient, tanaka1998mean}.
We give a simple derivation.

Fix \(h\) and consider unit \(i\).
For \(s_{-i}=(s_1,\ldots,s_{i-1},s_{i+1},\ldots,s_N)^\top\), define the conditional field
\[
H_i(s_{-i};h,K)
\coloneqq
h_i+\sum_{j\ne i}K_{ij}s_j .
\]
Because \(K=K^\top\) and \(K_{ii}=0\), 
\[
h^\top s+\frac12 s^\top Ks
=
s_i H_i(s_{-i};h,K)+C_i(s_{-i};h,K),
\]
where \(C_i\) does not depend on \(s_i\).
Hence, for \(s_i\in\{-1,+1\}\),
\[
P_{h,K}(S_i=s_i\mid S_{-i}=s_{-i})
=
\frac{\exp\{s_iH_i(s_{-i};h,K)\}}
{2\cosh H_i(s_{-i};h,K)} .
\]
Therefore,
\[
\mathbb E_{P_{h,K}}[S_i\mid S_{-i}=s_{-i}]
=
\tanh H_i(s_{-i};h,K).
\]
The exact mean satisfies
\[
m_i(h,K)
=
\mathbb E_{P_{h,K}}
\left[
\tanh\left\{
h_i+\sum_{j\ne i}K_{ij}S_j
\right\}
\right].
\]
The naive mean-field approximation ignores fluctuations of the random coupling contribution
\(\sum_{j\ne i}K_{ij}S_j\) and replaces it by its mean.
This gives the approximate self-consistency relation
\[
m_i(h,K)
\simeq
\tanh\left\{
h_i+\sum_{j\ne i}K_{ij}m_j(h,K)
\right\}.
\]
We define the mean-field approximation \(m^*(h,K)\) as a vector satisfying the corresponding fixed-point equation
\[
m_i^*(h,K)
=
\tanh\left\{
h_i+\sum_{j\ne i}K_{ij}m_j^*(h,K)
\right\}.
\]
Equivalently, 
\[
m^*(h,K)
=
\tanh\{h+Km^*(h,K)\},
\]
where \(\tanh\) is applied componentwise.
At the posterior field \(h(o)\), we write \(m^*(o)\coloneqq m^*(h(o),K)\), giving
\begin{equation}
    m^*(o)=\tanh\{h(o)+K m^*(o)\}.
    \label{eq:mf_fixed_point}
\end{equation}
The approximation is \(m^*(o)\approx m(o)\), not an exact identity.
Define
\[
R(o)=\operatorname{diag}\left(
1-m^*_1(o)^2,\ldots,1-m^*_N(o)^2
\right),
\qquad
D(o)=R(o)^{1/2} .
\]
Since $m_i^*(o)=\tanh\{h_i(o)+( Km^*(o))_i\}\in(-1,1)$, $1-m_i^*(o)^2>0$ for all $i$.
Thus $D(o)^{-1}$ exists.

To linearize the fixed point, perturb $h(o)$ by a small vector $\Delta h$ and the corresponding fixed point by $\Delta m^*$. 
Keeping only first-order terms in Eq.~\eqref{eq:mf_fixed_point} gives
\[
    \Delta m^*=R(o)\{\Delta h+K\Delta m^*\}.
\]
Equivalently,
\[
(I-R(o)K)\Delta m^*=R(o)\Delta h .
\]
The matrix
\[
    L(o)=R(o)K
\]
is the Jacobian of the mean-field mapping $f_o(x)= \tanh\{h(o)+Kx\}$ at $x = m^*(o)$. 
We then introduce a symmetric matrix
\[
M(o)=D(o)KD(o) .
\]
Since $K=K^\top$, $M(o)=M(o)^\top$. 
Moreover,
\[
    L(o)=D(o)M(o)D(o)^{-1},
\]
so $L(o)$ and $M(o)$ are similar matrices and have the same eigenvalues. 
Let
\[
    M(o)q_k(o)=\lambda_k(o)q_k(o),
    \qquad
    q_k(o)^\top q_\ell(o)=\delta_{k\ell}.
\]
We assume that the fixed point is stable, that is the spectral radius of the Jacobian is less than 1: 
\[
\rho(L(o)) = \rho(M(o)) \coloneqq \max_k |\lambda_k(o)| <1.
\]
In this case, $I-R(o)K = I-L(o)$ is regular.
The local mean-field susceptibility at the posterior state is
\begin{equation}
    \chi(o)
    \coloneqq\left.\frac{\partial m^*}{\partial h^\top}\right|_{h=h(o)}
    = (I-R(o)K)^{-1}R(o)
    =D(o)(I-M(o))^{-1}D(o) .
    \label{eq:susceptibility1}
\end{equation}
Using the completeness $I = \sum_k q_k(o) q_k(o)^\top$ and the spectral decomposition $M(o)=\sum_k \lambda_k q_k(o) q_k(o)^\top$, Eq.~\eqref{eq:susceptibility1} is
\begin{equation}
\chi(o) = D(o) \sum_k\left(\frac{1}{1-\lambda_k}q_k(o)q_k(o)^\top \right)D(o) = \sum_k\frac{D(o)q_k(o)q_k(o)^\top D(o)}{1-\lambda_k}.
\label{eq:susceptibility2}
\end{equation}
A positive near-critical mode is an eigenmode with $\lambda_k(o)$ close to $1$. 
In that case the factor $(1-\lambda_k(o))^{-1}$ in Eq.~\eqref{eq:susceptibility2} becomes large. 
Thus the symmetric matrix $M(o)=R(o)^{1/2}KR(o)^{1/2}$ defines the effective modes that control the local susceptibility.

\section*{Bayesian surprise in the effective-mode basis}
In this section, we derive a local expression for Bayesian surprise. 
Note that the expression analyzed below is not intended to reproduce the empirical variational-Bayes complexity term in the main text exactly. 
It captures the belief updating in an analytically tractable model.

For later calculations, we first examine the properties of the logarithm of the partition function $\psi(h) = \log Z(h, K)$ for a fixed $K$.
Here we define $m(h)$ as $m(h, K)$.
First, the derivative of $\psi(h)$ with respect to the field vector $h$ yields the exact mean $m$:
\begin{align*}
\frac{\partial\psi(h)}{\partial h_i}
&= \frac{1}{Z(h,K)}\frac{\partial Z(h,K)}{\partial h_i}
=\frac{1}{Z(h,K)}\sum_{s\in\Omega}\frac{\partial}{\partial h_i} \exp\left\{h^\top s+\frac{1}{2} s^\top K s\right\}\\
&= \frac{1}{Z(h,K)}\sum_{s\in\Omega}s_i\exp\left\{h^\top s+\frac{1}{2} s^\top K s\right\}\\
&= \sum_{s\in\Omega}s_i P_{h,K}(S=s) =m_i(h).
\end{align*}
Thus $\nabla_h \psi(h)=m(h)$.
Next, we show that the second derivative of $\psi(h)$ is the covariance and the Fisher information regarding $h$.
Since 
\begin{align*}
\nabla_h P_{h,K}(S=s)
&=\nabla_h \exp\left\{h^\top s+\frac{1}{2}s^\top K s-\psi(h)\right\}\\
&= P_{h,K}(S=s)\left(s-\nabla_h \psi(h)\right)\\
&= P_{h,K}(S=s)\left(s-m(h)\right),
\end{align*}
we obtain 
\begin{align*}
\frac{\partial^2\psi(h)}{\partial h_j \partial h_i} 
&= \frac{\partial m_i(h)}{\partial h_j}\\
&= \frac{\partial}{\partial h_j} \sum_{s\in\Omega} s_i P_{h,K}(S=s)\\
&=\sum_{s\in\Omega}s_i P_{h,K}(S=s) (s_j-m_j(h))\\
&= \sum_{s\in\Omega}s_is_j P_{h,K}(S=s)-m_j(h)\sum_{s\in\Omega}s_i P_{h,K}(S=s)\\
&= \mathbb{E}\left[S_i S_j\right] - \mathbb{E}\left[S_i\right]\mathbb{E}\left[S_j\right]\\
&= \mathrm{Cov}_{P_{h,K}}(S_i, S_j).
\end{align*}
Thus $\nabla_h^2\psi(h)=\partial m/\partial h^\top=\mathrm{Cov}_{P_{h,K}}\left[S\right]$.
Moreover this coincides with the Fisher information matrix with respect to $h$:
\begin{align*}
    F(h)
    &\coloneqq \mathbb{E}_{P_{h,K}}\left[\nabla_h\log P_{h,K}(S)\nabla_h\log P_{h,K}(S)^\top\right]\\
    &=\mathbb{E}_{P_{h,K}}\left[(S-m(h))(S-m(h))^\top\right]\\
    &=\mathrm{Cov}_{P_{h,K}}\left[S\right].
\end{align*}

We now derive the local expression for Bayesian surprise. 
Bayesian surprise is the KL divergence from the posterior belief to the prior belief:
\[
B(o)
=D_{\mathrm{KL}}(P_o\|P_p)
=\sum_{s\in\Omega}P_o(S=s)
\log\frac{P_o(S=s)}{P_p(S=s)} .
\]
The logarithm of the ratio of the probabilities is 
\[
\log \frac{P_o(S=s)}{P_p(S=s)}=(h(o)-h_p)^\top s-\psi(h(o))+\psi(h_p).
\]
Taking the expectation under the posterior distribution $P_o$,
\[
B(o)=(h(o)-h_p)^\top m(o)-\psi(h(o))+\psi(h_p).
\]

For a small field update $\delta h(o)=h(o)-h_p$ with $K$ fixed, the KL divergence has the second-order expansion as follows:
\begin{align*}
B(o)
&=\delta h(o)^\top m(h_p+\delta h(o))-\psi(h_p+\delta h(o))+\psi(h_p)\\
&= \delta h(o)^\top\left\{m(h_p)+F(h_p)\delta h(o) + O(\|\delta h(o)\|^2)\right\} \\
&- \left\{m(h_p)^\top \delta h(o) +\frac{1}{2} \delta h(o)^\top F(h_p) \delta h(o) + O(\|\delta h(o)\|^3)\right\}\\
&= \frac{1}{2}\delta h(o)^\top F(h_p)\delta h(o)+O(\|\delta h(o)\|^3).
\end{align*}
Equivalently, to second order one may use $F(h(o))=\operatorname{Cov}_{P_o}(S)$ instead of $F(h_p)$, because $F(h(o))-F(h_p)=F(h_p+\delta h(o))-F(h_p)=O(\|\delta h(o)\|)$ and therefore $\delta h(o)^\top F(h(o))\delta h(o) - \delta h(o)^\top F(h_p)\delta h(o) = O(\|\delta h(o)\|^3)$. 
We use $F(h(o))$ so that Bayesian surprise and the integrated-information proxy are evaluated at the same operating point.

Under the mean-field approximation, $F(h(o)) = \partial m/\partial h^\top\simeq \partial m^*/\partial h^\top = \chi(o)$. 
Combining this with Eq.~\eqref{eq:susceptibility2} gives
\[
    B(o)
    \simeq
    \frac{1}{2}\delta h(o)^\top \chi(o) \delta h(o)
    =
    \frac{1}{2}\sum_{k=1}^{N}
    \frac{
    \left|q_k(o)^\top D(o)\delta h(o)\right|^2
    }{
    1-\lambda_k(o)
    } .
\]
The numerator measures how strongly the observation-induced field update projects onto mode $k$, after weighted by $D(o)$. 
The denominator shows that this input projection is amplified when the same mode is positive near-critical.

\section*{Linear-Gaussian integrated-information proxy}

We now define a local integrated-information proxy.
The aim is to quantify how much the one-step local dynamics changes when couplings across a partition are removed, and then to express that quantity in the same eigenbasis of $M(o)$.
Note that the proxy analyzed here is not identical to the empirical \(\Phi_R^{\text{mc}}\), but is intended to capture the generic partition-sensitive integration.

A fixed observation $o$ specifies a mean-field fixed point $m^*(o)$ but not a unique temporal dynamics. 
To construct a local dynamical proxy for
integrated information, we introduce a time-dependent small fluctuation $z_t\in\mathbb R^N$ around the posterior mean-field fixed point $m^*(o)$.
Motivated by the linearization
\[m^*(o)+z_{t+1}=\tanh\{h(o)+K(m^*(o)+z_t)\} = m^*(o) + L(o)z_t +O(\|z_t\|^2),
\]
we use the local linear stochastic dynamics
\[
    z_{t+1}=L(o)z_t+\eta_t.
\]
In the transformed coordinate $y_t=D(o)^{-1}z_t$, this becomes
\[
    y_{t+1}=M(o)y_t+\xi_t,
    \qquad
    \xi_t = D(o)^{-1}\eta_t
    .
\]
Assume Gaussian noise with covariance $\operatorname{Cov}(\xi_t)=\sigma^2 I$.
Then the stationary mean is zero, and the stationary covariance $C(o)=\mathbb{E}[y_ty_t^\top]$ satisfies the Lyapunov equation
\[
    C(o)=M(o)C(o)M(o)^\top+\sigma^2 I.
\]
Since \(M(o)=M(o)^\top\) and \(\rho(M(o))<1\), the Lyapunov equation has the unique solution:
\[
C(o)=\sum_{k=1}^{N}\frac{\sigma^2}{1-\lambda_k(o)^2}q_k(o)q_k(o)^\top .
\]

We now derive the proxy for integrated information by considering how much the dynamics are disrupted when the connectivity $K$ is partitioned.
Let $\pi=\{V_1,\ldots,V_r\}$ be a disjoint and exhaustive partition of the units. 
Define $K_\pi$ by retaining within-partition couplings and removing cross-partition couplings:
\[
    (K_\pi)_{ij}=
    \begin{cases}
    K_{ij}, & i,j\in V_a\ \text{for some }a,\\
    0, & \text{otherwise}.
    \end{cases}
\]
At the same posterior operating point, define
\[
    M_\pi(o)=D(o)K_\pi D(o),
    \qquad
    \Delta_\pi M(o)=M(o)-M_\pi(o) .
\]
Here $R(o)$ and $D(o)$ are kept fixed. 
Thus the partition operation removes cross-partition couplings at the local posterior state; it does not recompute a new fixed point after the cut.

The integrated-information proxy for partition $\pi$ is the expected KL divergence between the full and partitioned one-step Gaussian transition distributions:
\begin{equation}
    \Phi_\pi(o)
    =\mathbb{E}_{y\sim\mathcal{N}(0,C(o))}
    \left[
    D_{\mathrm{KL}}\left(
    \mathcal{N}(M(o)y,\sigma^2 I)
    \middle\|
    \mathcal{N}(M_\pi(o)y,\sigma^2 I)
    \right)
    \right] .
    \label{eq:phi_def}
\end{equation}
For Gaussian distributions with the same covariance, Eq.~\eqref{eq:phi_def} becomes
\[
\Phi_\pi(o)
=\frac{1}{2\sigma^2}\operatorname{Tr}
\left(\Delta_\pi M(o) C(o)\Delta_\pi M(o)^\top\right)
=\frac{1}{2}\sum_{k=1}^{N}
\frac{
\|\Delta_\pi M(o)q_k(o)\|^2
}{
1-\lambda_k(o)^2
} .
\]
Thus mode $k$ contributes strongly to $\Phi_\pi(o)$ when the partition substantially destroys that mode.

Here, for simplicity, we use the atomic partition \(\pi=\{\{1\},\{2\},\ldots,\{N\}\}\).
Since $\Delta_\pi M(o)=M(o)$, we obtain the following form:
\[
\Phi_{\operatorname{atom}}(o)=\frac12\sum_k \frac{\lambda_k(o)^2}{1-\lambda_k(o)^2}.
\]

\section*{Shared amplification under positive near-critical modes}

Having written Bayesian surprise and the integrated-information proxy in the same eigenbasis, we extract the main theoretical result. 
The goal is to show that a single positive near-critical mode can simultaneously amplify belief updating and partition-sensitive integration. 
This is the mode-level mechanism that supports the empirical observation that integrated information covaries robustly with Bayesian surprise.

The Bayesian surprise and integrated information are locally approximated as:
\[
B(o)\simeq
\frac{1}{2}\sum_k\frac{A_k(o)}{1-\lambda_k(o)},
\qquad
\Phi_{\operatorname{atom}}(o)\simeq
\frac{1}{2}\sum_k\frac{\lambda_k(o)^2}{1-\lambda_k(o)^2},
\]
where \(A_k(o)=|q_k(o)^\top D(o)\delta h(o)|^2\).

Suppose one positive critical mode $\lambda_\ast(o)\to1$. 
Since
\[
    1-\lambda_\ast(o)^2
    =(1-\lambda_\ast(o))(1+\lambda_\ast(o))
    \simeq 2(1-\lambda_\ast(o)),
\]
we obtain
\[
B(o)
\simeq
\frac{A_\ast(o)}{2(1-\lambda_\ast(o))},
\qquad
\Phi_{\operatorname{atom}}(o)
\simeq
\frac{1}{4(1-\lambda_\ast(o))},
\qquad
\therefore B(o)\simeq2A_\ast(o)\Phi_{\operatorname{atom}}(o)
\]
Both quantities are therefore amplified by the same effective critical factor, $(1-\lambda_\ast(o))^{-1}$.
In this regime, informative inputs that project onto a critical mode increase Bayesian surprise.
For the atomic partition, or more generally for partitions that substantially disrupt the same mode, the integrated-information proxy is amplified by the same critical factor.

\paragraph{Conditions for correlation between Bayesian surprise and integrated
information.}
The approximations for \(B(o)\) and \(\Phi_\mathrm{atom}(o)\) show that both quantities depend on the eigenvalues of \(M(o)\).
Their correlation arises under the following conditions:
\begin{itemize}
    \item The dominant eigenvalue \(\lambda_\ast\) is close to \(1\), indicating a critical mode.
    \item \(\lambda_\ast\) is positive. If \(\lambda_\ast\to-1\), \(\Phi_\mathrm{atom}(o)\) can diverge while \(B(o)\) remains finite, weakening the correlation. 
    \item The observation-induced change in local field \(\delta h\) has a non‑zero projection onto the eigenvectors \(q_k\), so that the numerator of \(B(o)\) in the approximations does not vanish.
    \item The variation in the shared critical factor dominates over variation in the numerator \(A_\ast(o)\).
\end{itemize}

\section*{Hill-shaped trajectory of $\Phi$ through learning}
The previous sections considered a fixed interaction matrix \(K\). 
We suppose the interaction matrix $K$ slowly changes through learning and show that \(\Phi\) can follow a non-monotonic hill-shaped trajectory.
We denote the learning stage by \(\ell\) and write

\[
M_\ell(o)=D_\ell(o)K_\ell D_\ell(o),
\qquad
D_\ell(o)=R_\ell(o)^{1/2},
\qquad
R_{\ell,i}(o)=1-m_{\ell,i}^*(o)^2.
\]

Let \(\lambda_\ast\) be a simple dominant positive eigenvalue of \(M_\ell(o)\), with normalized eigenvector \(q_\ast\). 
Differentiating
\[
M_\ell=D_\ell K_\ell D_\ell
\]
with respect to learning stage \(\ell\) gives

\[
\frac{d\lambda_\ast}{d\ell}
=
q_\ast^\top
\frac{dM_\ell}{d\ell}
q_\ast.
\]

For notational simplicity, we suppress the dependence on \(\ell\) and \(o\) in the following expression.
Using \(D_i=(1-{m_i^*}^2)^{1/2}\), this becomes

\begin{equation}
\frac{d\lambda_\ast}{d\ell}
=
(Dq_\ast)^\top
\frac{dK}{d\ell}
(Dq_\ast)
-
2\lambda_\ast
\sum_i
\frac{m_i^*\,dm_i^*/d\ell}{1-{m_i^*}^2}
q_{\ast i}^2
\label{eq:eigenvalue_derivation}
\end{equation}

This decomposition gives a simple interpretation.
The first term increases \(\lambda_\ast\) when learning strengthens the relevant effective interaction connectivity. 
This does not mean that all entries of \(K\) are positive or increase elementwise; \(K\) may be signed. 
It only means that the positive collective interaction mode relevant for \(M_\ell\) becomes stronger.
The second term decreases \(\lambda_\ast\) when the response $m_i^*$ approaches deterministic values, $\pm 1$.

We now provide a possible scenario during learning.
Early in learning, the posterior means are close to neutral zero, $m_i^*\simeq 0$, because of a lack of evidence.
$R_i=1-{m_i^*}^2\simeq 1$ and the second term of \eqref{eq:eigenvalue_derivation} is small. 
In this regime, strengthening of interaction connectivity $K$ directly increases the eigenvalue \(\lambda_\ast\), and therefore increases \(\Phi_{\mathrm{atom}}\).

Later in learning, the response may become stabilized. 
In the binary idealization, this corresponds to \(m_i^*\) moving away from \(0\) and toward \(\pm1\), and the local susceptibility $R_i=1-{m_i^*}^2$ decreases. 
In the stabilization phase, learning tends to push each posterior mean toward its selected binary state, \(m_i^*>0\) toward \(+1\) and \(m_i^*<0\) toward \(-1\), implying \(m_i^*\, dm_i^*/d\ell\ > 0\).
Thus the second term can dominate the first term and reduce \(\lambda_\ast\). 
Since \(\Phi_{\mathrm{atom}}\) increases with positive \(\lambda_\ast\), this produces a rise-and-fall trajectory of integrated information.

This mechanism does not imply that \(\Phi_{\mathrm{atom}}\) decreases forever. 
Once learning has converged, both \(dK/d\ell\) and \(dm^*/d\ell\) approach zero, so \(d\lambda_\ast/d\ell\) also approaches zero. 
The expected trajectory is therefore a rise, a peak, and then a lower plateau.

A minimal one-dimensional mean-field model simply illustrates this idea. 
Consider
\[
m=\tanh(Jm),
\qquad
R=1-m^2,
\qquad
\lambda=JR,
\]
where \(J\) represents the strength of the connectivity, and \(R\) is the local susceptibility.
This one-dimensional construction should not be interpreted as a literal $N=1$ case. 
It is only a schematic example showing how the effective eigenvalue can increase with coupling strength $J$, but decrease when local susceptibility $R=1-m^2$ is reduced.

Consider the change in the eigenvalue $\lambda$ as the strength \(J\) changes. 
The derivative is given by the following:
\[
\frac{d\lambda}{dJ}
=R+J\frac{dR}{dJ}
=
R
-
2Jm\frac{dm}{dJ}.
\]
For \(dm/dJ\), we obtain an expression as follows:
\[
\frac{dm}{dJ}=\frac{d}{dJ}\tanh(Jm)=\left(1-\tanh^2(Jm)\right)\left(m+J\frac{dm}{dJ}\right)=R\left(m+J\frac{dm}{dJ}\right), 
\qquad
\therefore \frac{dm}{dJ}=\frac{Rm}{1-\lambda}.
\]

We now show that \(\lambda\) has a peak.
For \(J<1\), the stable solution is \(m=0\), so \(R=1\) and \(d\lambda/dJ=1\).
Thus, in this regime, increasing the strength \(J\) directly increases the effective eigenvalue. 
At \(J=1\), the system reaches the mean-field phase transition. 
For \(J>1\), the stable branch becomes ordered, \(m\neq0\). 
Along this branch,
\[
\frac{d\lambda}{dJ}
=
R\left(
1-\frac{2Jm^2}{1-\lambda}
\right) = -R\frac{(J-1)+Jm^2}{1-\lambda}
<0,
\]
where \(1-\lambda>0\) because the ordered branch is locally stable.
Thus, once the response becomes ordered, the reduction of local susceptibility \(R=1-m^2\) dominates the direct increase of \(J\), pushing the effective eigenvalue \(\lambda=JR\) back down.

Thus the eigenvalue rises up to the transition and then decreases on the ordered branch. 
Since the integrated information here \(\Phi_{\mathrm{atom}}=\lambda^2/\{2(1-\lambda^2)\}\) is monotone increasing for \(0<\lambda<1\)
, the proxy peaks around the transition. 
In finite, noisy, or heterogeneous systems, this singular transition would appear as a rounded hill-shaped trajectory rather than a sharp divergence.

Note that this one-dimensional model is only an intuitive picture and the empirical analysis does not claim that the neural cultures underwent an exact mean-field phase transition.

\paragraph{Conditions for the hill-shaped trajectory of \(\Phi\).}
\begin{itemize}
    \item The dominant eigenvalue \(\lambda_\ast\) is close to \(1\), indicating a positive critical mode.
    \item Early in learning, the posterior mean is neutral: \(m_i^*\simeq 0\).
    \item Early learning strengthens the interaction connectivity \(K\).
    \item Later learning stabilizes the response and reduces the local susceptibility: \(m_i^*\simeq\pm1\) and \(R_i\simeq0\).
\end{itemize}

\section*{Summary}

The theoretical analysis presented here gives a possible mechanism for why Bayesian surprise and integrated information covary, and integrated information follows a non-monotonic hill-shaped trajectory in the experiments. 
Both quantities are influenced by the system’s effective critical modes; near a positive critical mode, they are amplified together because their denominators contain \(1-\lambda_k\) and \(1-\lambda_k^2\).
The relationship is not trivial or guaranteed: a dominant positive critical mode must exist, and the input-driven field change must project strongly onto that critical mode. 
Furthermore, the derivative of the critical eigenvalue with respect to the learning stage decomposes into two terms representing an increase in effective coupling strength and a decrease in sensitivity associated with more deterministic responses.
Under the assumption that the posterior response that was neutral in the early stages of learning becomes deterministic in the later stages as the coupling strength increases, this formulation provides a possible scenario for the hill-shaped trajectory of \(\Phi\).
These insights provide one possible mechanistic connection between the Bayesian perspective of the FEP and the integration‑focused view of IIT, complementing the empirical results discussed in the main text.

\bibliographystyle{unsrt}
\bibliography{supplementary/si-references}